\documentclass[fleqn,usenatbib]{mnras}

\usepackage{float}
\usepackage{multirow}
\usepackage{epsfig}
\usepackage{array}
\usepackage{graphicx}
\usepackage{graphics}
\usepackage{subfigure}
\usepackage{latexsym}
\usepackage{amsmath}

\usepackage[T1]{fontenc}

\DeclareRobustCommand{\VAN}[3]{#2}
\let\VANthebibliography\thebibliography
\def\thebibliography{\DeclareRobustCommand{\VAN}[3]{##3}\VANthebibliography}

\usepackage{graphicx}	
\usepackage{amsmath}	
\usepackage{amssymb}	

\title[Optical and UV Monitoring of MAXI J1820+070]{Optical and Ultraviolet Monitoring of the Black Hole X-ray Binary MAXI J1820+070/ASASSN-18ey for 18 Months}

\author[Hanna Sai et al.]{
Hanna Sai,$^{1}$\thanks{E-mail: shn17@mails.tsinghua.edu.cn}
Xiaofeng Wang,$^{1,2}$\thanks{E-mail:wang\_xf@mail.tsinghua.edu.cn}
Jianfeng Wu,$^{3}$
Jie Lin,$^{1}$
Hua Feng,$^{4,5}$
Tianmeng Zhang,$^{6,7}$
\newauthor
Wenxiong Li,$^{1}$
Jujia Zhang,$^{8,9,10}$
Jun Mo,$^{1}$
Tianrui Sun,$^{11,12,13}$
Shuhrat A. Ehgamberdiev,$^{14}$
\newauthor
Davron Mirzaqulov,$^{14}$
Liming Rui,$^{1}$
Weili Lin,$^{1}$
Xulin Zhao,$^{15}$
Han Lin,$^{1}$
Jicheng Zhang,$^{1}$
\newauthor
Xinghan Zhang,$^{1}$
Yong Zhao,$^{6,16}$
Xue Li,$^{1}$
Danfeng Xiang,$^{1}$
Lingzhi Wang,$^{17,18}$
Chengyuan Wu$^{1}$
\\
$^{1}$Physics Department and Tsinghua Center for Astrophysics (THCA), Tsinghua University, Beijing, 100084, China\\
$^{2}$Beijing Planetarium, Beijing Academy of Science and Technology, Beijing, 100044, China\\
$^{3}$Department of Astronomy Xiamen University (Haiyun Campus), Siming District, Xiamen, Fujian, 361005, China\\
$^{4}$Department of Astronomy, Tsinghua University, Beijing, 100084, China. \\
$^{5}$Department of Engineering Physics, Tsinghua University, Beijing, 100084, China.\\
$^{6}$Key Laboratory of Optical Astronomy, National Astronomical Observatories, Chinese Academy of Sciences, Beijing, 100101, China\\
$^{7}$School of Astronomy and Space Science, University of Chinese Academy of Sciences, 101408, Beijing\\
$^{8}$Yunnan Observatories (YNAO), Chinese Academy of Sciences, Kunming 650216, China\\
$^{9}$Key Laboratory for the Structure and Evolution of Celestial Objects, Chinese Academy of Sciences, Kunming 650216, China\\
$^{10}$Center for Astronomical Mega-Science, Chinese Academy of Sciences, 20A Datun Road, Chaoyang District, Beijing, 100012, China\\
$^{11}$Purple Mountain Observatory, Chinese Academy of Sciences, Nanjing 210008, China\\
$^{12}$Chinese Center for Antarctic Astronomy, Nanjing 210008, China\\
$^{13}$School of Astronomy and Space Science, University of Science and Technology of China, Hefei 230026, China\\
$^{14}$Ulugh Beg Astronomical Institute, Uzbekistan Academy of Sciences, Uzbekistan, Tashkent, 100052, Uzbekistan\\
$^{15}$School of Science, Tianjin University of Technology, Tianjin, 300384, China\\
$^{16}$University of Chinese Academy of Sciences, Beijing 100049, China\\
$^{17}$National Astronomical Observatory of China, Chinese Academy of Sciences, Beijing, 100012, China\\
$^{18}$Chinese Academy of Sciences South America Center for Astronomy, National Astronomical Observatories, CAS, Beijing 100101, China
}

\date{Accepted XXX. Received YYY; in original form ZZZ}

\pubyear{2020}

\begin{document}
\label{firstpage}
\pagerange{\pageref{firstpage}--\pageref{lastpage}}
\maketitle

\begin{abstract}

MAXI J1820+070 is a low-mass black hole X-ray binary system with high luminosity in both optical and X-ray bands during the outburst periods. We present extensive photometry in X-ray, ultraviolet, and optical bands, as well as densely-cadenced optical spectra, covering the phase from the beginning of optical outburst to $\sim$550 days. During the rebrightening process, the optical emission preceded the X-ray by 20.80 $\pm$ 2.85 days. The spectra are characterized by blue continua and emission features of Balmer series, He~{\sc i}, He~{\sc ii} lines and broad Bowen blend. The pseudo equivalent width (pEW) of emission lines are found to show anticorrelations with the X-ray flux measured at comparable phases, which is due to the increased suppression by the optical continuum. At around the X-ray peak, the full width at half maximums (FWHMs) of H$_{\beta}$ and He~{\sc ii} $\lambda$4686 tend to stabilize at 19.4 Angstrom and 21.8 Angstrom,  which corresponds to the line forming region at a radius of 1.7 and 1.3 $R_{\sun}$ within the disk. We further analyzed the absolute fluxes of the lines and found that the fluxes of  H$_{\beta}$ and He~{\sc ii} $\lambda$4686 show positive correlations with the X-ray flux, favoring that the irradiation model is responsible for the optical emission. However, the fact that X-ray emission experiences a dramatic flux drop at t$\sim$200 days after the outburst, while the optical flux only shows little variations suggests that additional energy such as viscous energy may contribute to the optical radiation in addition to the X-ray irradiation.
\end{abstract}



\section{Introduction} \label{sec:intro}
Low-mass X-ray binaries (LMXBs) are close binary systems consisting of a compact object and a low-mass companion star having filled its Roche lobe. The mass-donor companion can be a main-sequence star, a subgiant, or a white dwarf (WD), while the compact object can be a black hole (BH) or a neutron star (NS). Materials of the donor star transfer to the compact object through the $L1$ Lagrangian point, forming a surrounding accretion disc \citep{2006csxs.book..215C, 2019ApJ...872L..20G}. LMXBs often stay for years or decades in a state of quiescence with X-ray luminosities of $\sim10^{29} - 10^{33.5}$erg s$^{-1}$ before turning into outbursts \citep{2006ARA&A..44...49R}. The mass of a BH in an X-ray binary can be determined based on the periodic variations inferred from the spectroscopic and photometric observations in quiescence (e.g.,\citealt{2014SSRv..183..223C, 2015ApJ...806...92W, 2016ApJ...825...46W}). During the outburst, the luminosity of these LMXBs can increase by many orders of magnitudes relative to its quiescent state \citep{1997ApJ...491..312C}. The outbursts can be explained by instability of accretion discs surrounding the compact object \citep{1995PASJ...47...47O, 2001NewAR..45..449L, 2001A&A...373..251D}, which is also called the ``thermal-viscous disc instability model" (DIM). Some LMXB transients can experience several outbursts, and those with higher mass transfer rate tend to have higher outburst frequencies \citep{2019ApJ...870..126L}.

According to the irradiated disk model, the optical and ultraviolet (UV) -band emissions are believed to arise from reprocessing of the X-ray emission \citep{2007ApJ...666.1129R,2008MNRAS.388..753G, 2009MNRAS.392.1106G}. In this scenario, the temperature structure of the outer disc will be altered by the irradiated inner disk and Compton tail, leading to production of the observed optical/UV emission \citep{2009MNRAS.392.1106G}. The optical spectra of LMXBs often contain emission lines, such as the Balmer series. The emission lines are believed to originate from rotating accretion flow of the disc, which is ionized by irradiation from a central high-luminosity X-ray source. Note, however, that there are other arguments about the origin of the optical radiation of LMXB transients, such as the jet/corona \citep{1998ApJ...506L..31E,2005ApJ...635.1203M,2006MNRAS.371.1334R} or the atmosphere of an optically thick disc in bright/soft X-ray states \citep{2001MNRAS.320..177W, 2002nmgm.meet.2274W}. Moreover, the H$_{\alpha}$ emission line could also originate in a dense outflow in low/hard X-ray states \citep{2001MNRAS.320..177W}. At low-luminosity state, the ionizing source is usually considered to be viscous heating of the disc \citep{1994A&A...290..133V, 2009MNRAS.393.1608F}. 

The outburst of the LMXB system, MAXI J1820+070 (ASASSN-18ey), was initially discovered as an optical transient ASASSN-18ey on UT Mar.06.58 2018 by the All Sky Automated Survey for SuperNovae (ASAS-SN; \citealt{2014ApJ...788...48S, 2017PASP..129j4502K}) at R.A. = $18^h20^m21^s.9 $, dec. = $ +07^{\circ}11'07''.3$ (J2000) \footnote{http://www.astronomy.ohio-state.edu/asassn/transients.html}. About one week later, the Monitor of All-sky X-ray Image (MAXI; \citealt{2009PASJ...61..999M}) Gas Slit Camera \citep{2011PASJ...63S.623M} nova alert system detected a bright X-ray transient at the same location with a flux of 32 $\pm$ 9 mCrab in 4 - 10 keV \citep{2018ATel11399....1K}, which was designated as MAXI J1820+070. Multiwavelength follow-up observations suggested that MAXI J1820+070 is a BH LMXB \citep[e.g.,][]{2018ATel11418....1B, 2018ATel11820....1H, 2018ApJ...867L...9T}. The distance to this system is estimated as $3.46^{+2.18}_{-1.03}$~kpc according to the Gaia DR2 parallax data \citep{2019MNRAS.485.2642G}. \cite{2020MNRAS.493L..81A} provided a consistent and more precise measurement on the distance to MAXI J1820+070 as 2.96$\pm$0.33~kpc using the parallax obtained from radio interferometry.

The X-ray and optical monitoring of this system showed that there is a time lag of 7.20$\pm$0.97 days between the outbursts detected in optical and X-ray bands for MAXI J1820+070 in March, 2018 \citep{2018ApJ...867L...9T}. This delay suggests that the thermal instability initially triggered an outburst in the thermal-viscous disc. The X-ray and optical monitoring of this system suggests that the state transition is not only related to the mass accretion rate \citep{2019ApJ...874..183S}. They proposed that the jet contributed to the optical emission in the low/hard state, whereas the outer disc emission dominated the optical flux in high/soft state. Polarization observations also favored for the existence of jet or hot flow for MAXI J1820+070 \citep{2019A&A...623A..75V}. Dynamical modeling of MAXI J1820+070 system was carried out by \cite{2019ApJ...882L..21T}. They confirmed the property of stellar-mass BH for the compact primary of MAXI J1820+070, with its mass function as $f(M)=5.18 \pm 0.15 M_{\sun}$ and an orbital period of $0.68549 \pm 0.00001$ days. \cite{2019ApJ...879L...4M} detected P-Cygni profiles and broad wings in He~{\sc i} $\lambda$5876 and H$_{\alpha}$ emission lines in the high-resolution optical spectra of MAXI J1820+070 taken during low/hard states, suggesting the existence of accretion disc wind.

In this paper, we report extensive follow-up observations of MAXI J1820+070 in optical, UV, and X-ray bands for about 18 months, and present its observational properties. The relations between optical/UV and X-ray features have been analyzed, with an attempt to better constrain the radiation physics for this BH binary system. The paper is organized as follows: in Section~\ref{sec:obser}, the observations and data reductions are described. Section~\ref{sec:lc} presents the light/color curves, and Section~\ref{sec:spectra} presents the spectral evolution. In Section~\ref{sec:dis}, we discuss the properties of MAXI J1820+070, including spectral energy distribution (SED) and the correlation between optical and X-ray properties. We summarize our results in Section~\ref{sec:con}.

\section{Observations and Data Reduction} \label{sec:obser}

\begin{figure}
\includegraphics[width=\columnwidth]{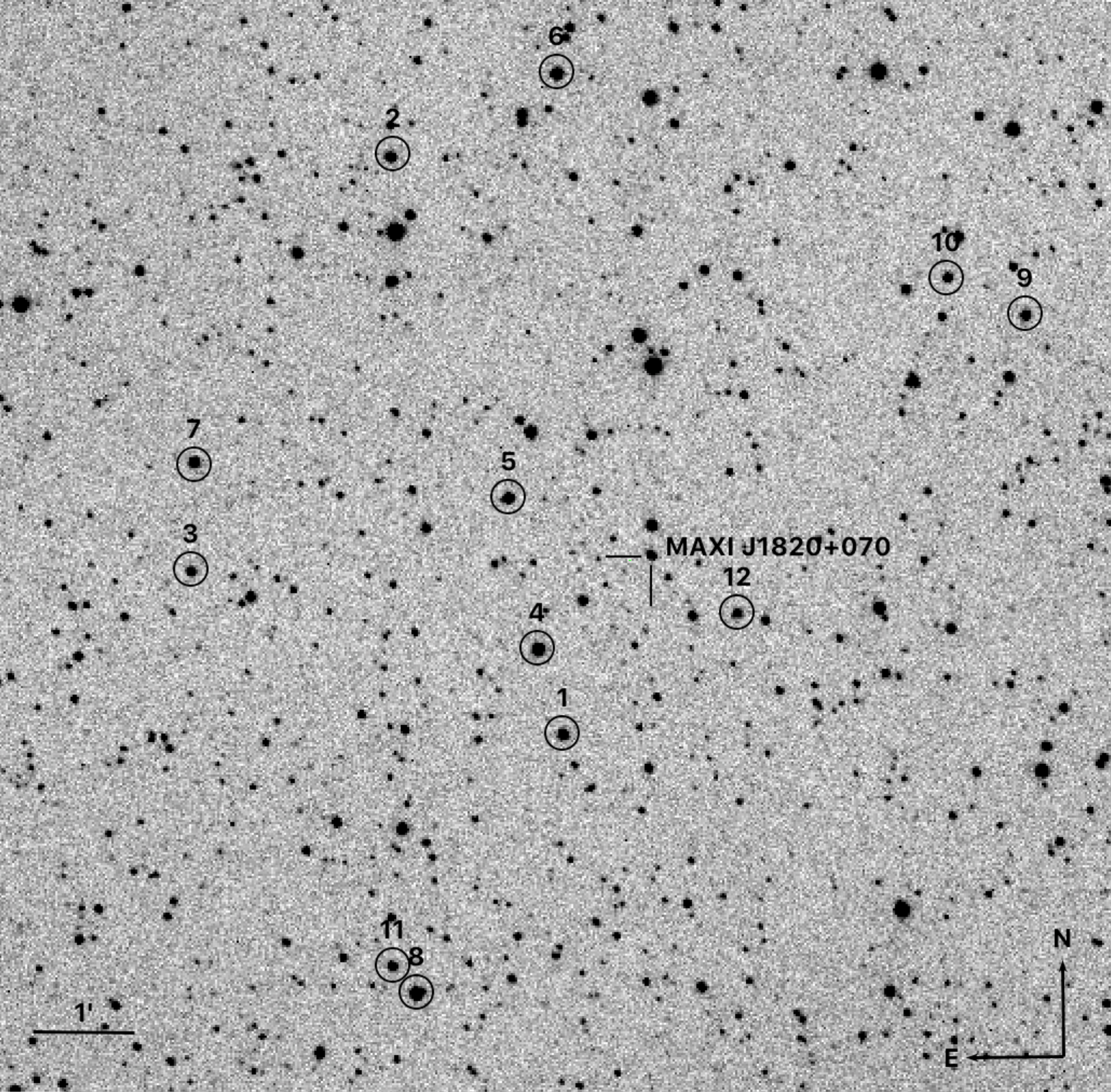}
\caption{Image of MAXI J1820+070/ASASSN-18ey, taken with the Tsinghua-NAOC 0.8-m Telescope (TNT). North is up and east is to the left. MAXI J1820+070 is indicated by the solid thick sticks. 
The reference stars are labeled with black circles.}
\label{verify picture}     
\end{figure}

After the discovery of MAXI J1820+070, we started a long-term spectroscopic and photometric monitoring campaign in optical bands. The follow-up photometric observations started on March 20, 2018, using the 0.8-m Tsinghua-NAOC Telescope (TNT) (\citealt{2012RAA....12.1585H}; see Figure~\ref{verify picture} for a sample image) and the Yaoan High Precision Telescope in China, as well as the AZT-22 1.5-m telescope (hereafter AZT) at the Maidanak Astronomical Observatory in Uzbekistan \citep{2018NatAs...2..349E}. We use standard IRAF routines to pre-process all CCD images, which include bias subtraction, flat fielding, and the removal of cosmic rays. Point-spread-function (PSF) photometry was performed for both the object and the reference stars using the pipeline $Zuruphot$ developed for automatic photometry of TNT (Mo et al. in prep.). The instrumental magnitudes were then converted into those of the Johnsons $BV$ \citep{1966CoLPL...4...99J} and Sloan Digital Sky Survey (SDSS) $gri$-band system \citep{1996AJ....111.1748F}. Table~\ref{standard} lists the standard BV- and $gri$-band magnitudes of the comparison stars, which are also marked in Figure~\ref{verify picture}. MAXI J1820+070 was also observed by the Ultraviolet/Optical Telescope (UVOT; \citealt{2005SSRv..120...95R}) onboard the \emph{Neil Gehrels Swift Observatory} (\emph{Swift}; \citealt{2004ApJ...611.1005G}) in three UV ($uvw2$, $uvm2$ and $uvw1$) and three optical filters ($u$, $b$ and $v$). The final calibrated magnitudes in the \emph{Swift} filters are presented in Table~\ref{table_lc} and Table~\ref{table_lc_uv}, respectively. The \emph{Swift} images of MAXI J1820+070 were reduced using the HEASOFT \footnote{HEASOFT, the High Energy Astrophysics Software https://www.swift.ac.uk/analysis/software.php} with the latest \emph{Swift} calibration database \footnote{https://heasarc.gsfc.nasa.gov/docs/heasarc/caldb/swift/}.

We also analyzed the \emph{Swift}/XRT observations of MAXI J1820+070 obtained during the period from MJD 58189 to MJD 58591. These observations were obtained in both window timing (WT) and photon count (PC) modes. All of the XRT data were first reduced by running the $xrtpipeline$ task of HEASOFT (version 6.19). Since the count rate of MAXI J1820+070 can reach beyond 500 cps, the source region suffers serious pile-up effect. To avoid this effect, we extract the source photons using an annular region. The outer radius is set as 40 pixels, and the inner radius is adjustable according to the count rates. In the WT mode observations, the inner radius is set as RATE/100 pixels (RATE stands for the count rate). In the PC mode observations, the inner radius is fixed as 2 pixels if the count rate is higher than 0.5 cps; otherwise, the inner radius is set to zero. The background photons were estimated from an annular region with an outer radius of 120 pixels and an inner radius of 60 pixels. The energy spectra were extracted by the XSELECT package and grouped into bins with a minimum of 20 photons. We adopted response matrix files (RMFs) from the \emph{Swift}/XRT calibration database (CALDB) and created ancillary response files (ARFs) from exposure maps using the $xrtmkarf$ task. The XRT spectra between 0.6 and 10 keV were fitted with a model composed of a power-law component (the $powerlaw$ model) plus a multi-black body component (the $diskbb$ model). The power-law component accounts for the comptonized X-ray emission, and the multi-black body component represents the thermal X-ray emission from the surface of the accretion disc, while the X-ray absorption is characterized by the $wabs$ model. During the phase from 336 to 371 days after the initial detection, the X-ray count rate is quite low so we use a single power-law model to estimate the flux \citep{2005SPIE.5898...22K}. 

A total of 66 spectra were obtained for MAXI J1820+070 with the Lijiang 2.4-m telescope (+YFOSC) and Xinglong 2.16-m telescope (+BFOSC and +OMR) \citep{2016PASP..128j5004Z}, covering the phases from +11.9 days to +385.3 days since the discovery. A log of the spectra is given in Table~\ref{table_log}. We reduced all the spectra using standard IRAF routine. Flux calibration of the spectra was performed with spectrophotometric standard stars taken on the same nights. The spectra were corrected for atmospheric extinction using the extinction curves of local observatories; and the telluric lines were removed from the spectra.

\section{Light Curves} \label{sec:lc}
\begin{figure*}
\centering
\includegraphics[scale=0.65]{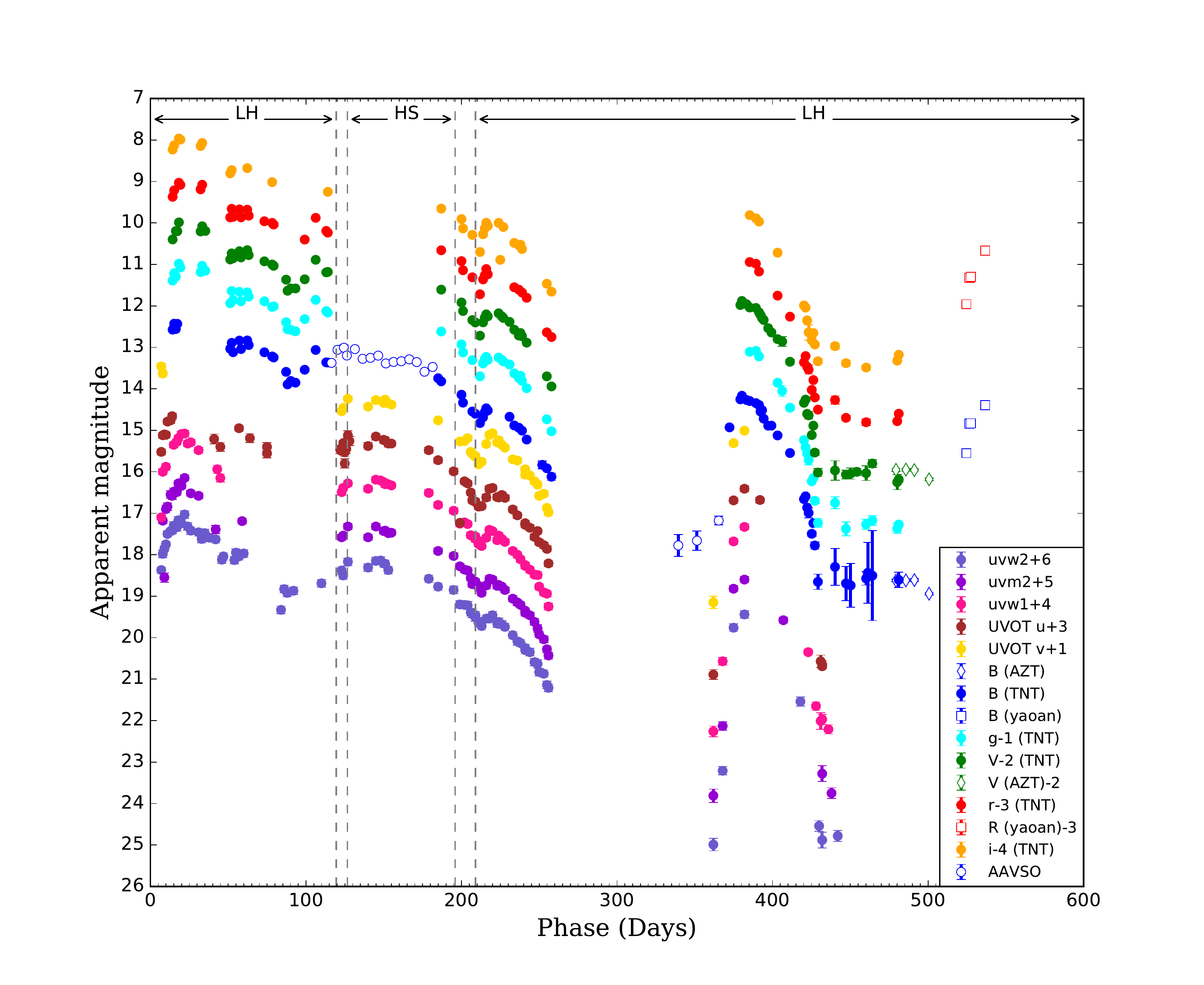}
\caption{The optical and UV light curves obtained for MAXI J1820+070, lasting for about 550 days after the initial optical outburst (UT March 06.58 2018 = MJD 58184.08). Different colors represent different bands, including the \emph{Swift} $uvw2$, $uvm2$, $uvw1$, UVOT $u$, UVOT $v$, $B$, $g$, $V$, $r$, $R$, $i$, and $I$ bands. The light curves of different bands have been vertically shifted for better display. The arrow labels with `LH'  and `HS' on the top represent the low/hard and high/soft state periods, respectively. The two narrow gaps between the LH and HS are intermediate-state periods \citep{2019ApJ...874..183S}.} 
\label{long_lc}     
\end{figure*}

Figure~\ref{long_lc} shows the optical and UV light curves of MAXI J1820+070, covering the phase from +6.92 to +536.67 days relative to the first detection of optical outburst. For the TNT observations, the first gap appeared between July and August, 2018, covering the phase from +114 to +185 days, is due to the maintenance of the telescope in the summer season, while the second gap, seen during the phase from t$\sim$ +258 to +372 days, is because that the source moved close to the sun. To complement the sampling of the light curves, we also collected the $B$-band data from the American Association of Variable Star Observers (AAVSO). 

At t $\sim$ +100 days and 210 days after outburst, MAXI J1820+070 is found to exhibit two rebrightening behaviors. However, it did not show quiescence between these two epochs. The first rebrightening occurred before the transition from low/hard to high/soft states, while the second one occurred shortly after the transition from the high/soft to low/hard states. By t$\sim$ +360 days, this  system faded gradually towards its quiescent state \citep{2019ATel12534....1R}. At t $\sim$ +370 days, however, this system is found to experience a post-outburst ``rebrightening" in both optical and X-ray bands \citep{2019ATel12567....1U,2019ATel12573....1B},  and reached a B-band peak of 14.17 mag at t$\sim$+380 days. This corresponds to an absolute magnitude of 1.47 mag in B band using the Gaia distance \citep{2019MNRAS.485.2642G}. After that, MAXI J1820+070 tended to decrease in both X-ray and optical bands and entered into the quiescent state based on the X-ray flux reported by \emph{Swift} and \emph{NuSTAR} \citep{2019ATel12688....1V,2019ATel12732....1T}. At t$\sim$+507 days, MAXI J1820+070 was found to rebrighten again in opitcal (see Figure~\ref{long_lc}) and other bands including radio, near-infrared and X-ray bands (see \citealt{2019ATel13014....1H}, \citealt{2019ATel13025....1X}, \citealt{2019ATel13041....1B} and \citealt{2019ATel13044....1H}).

\section{Optical Spectra} \label{sec:spectra}
\subsection{Spectral Evolution}
\begin{figure*}
  \subfigure{
  \label{spec_cor1}
   \includegraphics[scale=.5]{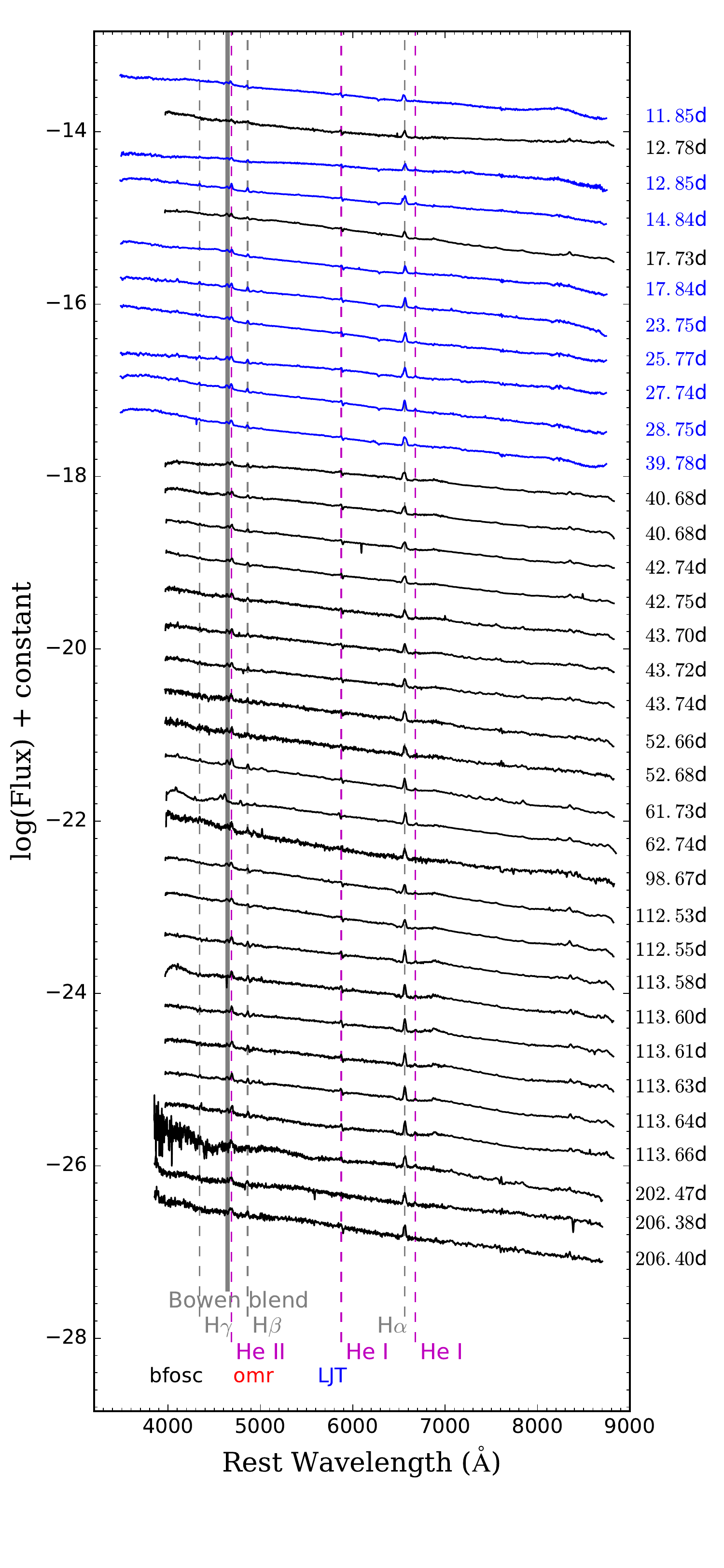}}
  \subfigure{
     \label{spec_cor2} 
   \includegraphics[scale=.5]{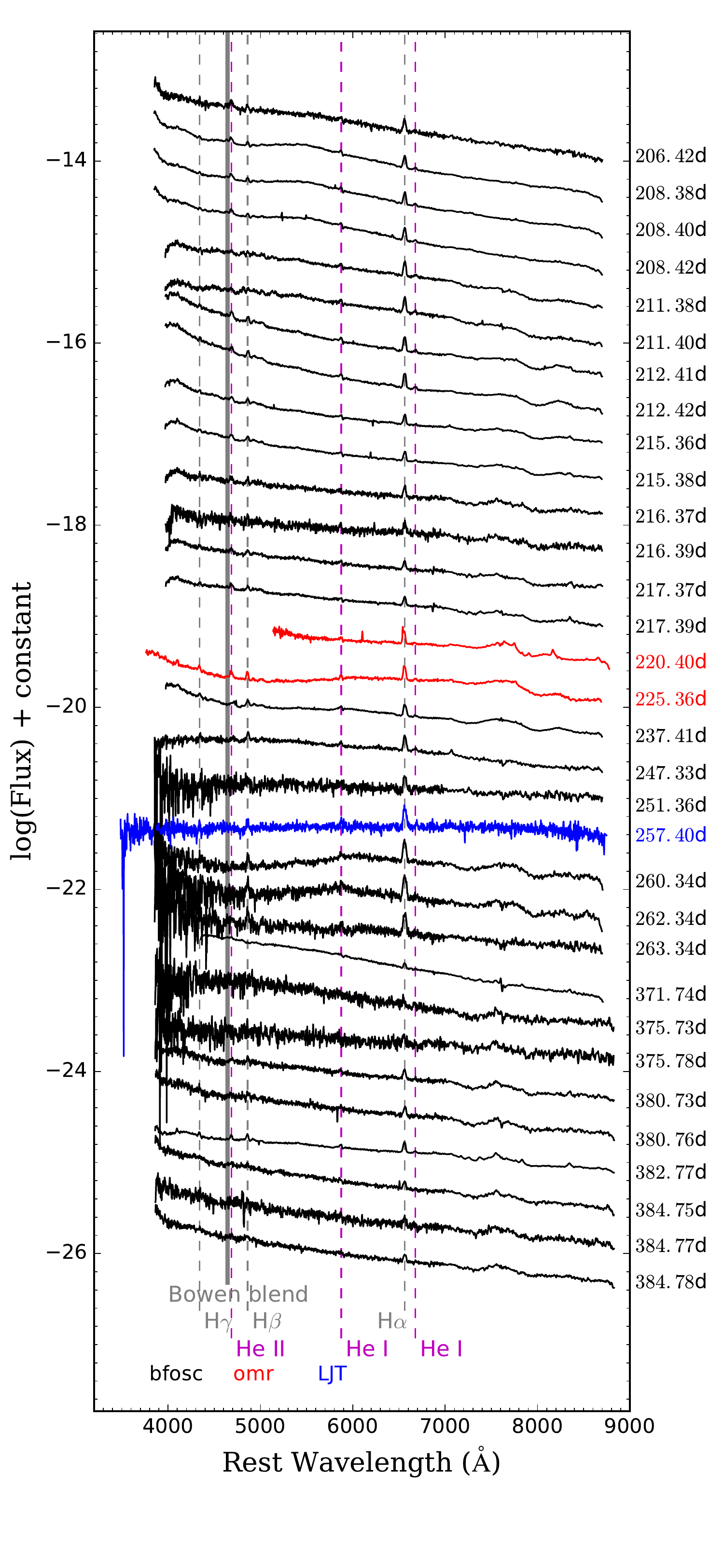}}
      \caption{Spectral evolution of MAXI J1820+070. The continuums of the spectra have been flux-calibrated by the photometry obtained at similar phases. The spectra have been shifted vertically for better display. The epoch on the right side of each spectrum represents the phase in days after the initial optical outburst. The Balmer series (including H$_{\alpha}$, H$_{\beta}$, and H$_{\gamma}$ here) and the helium emission lines are marked by grey and magenta dashed lines, respectively. The Bowen blend feature is labeled by the grey vertical bar in each panel. Different colors of the spectra represent different spectroscopic instruments (i.e., XLT+bfosc, XLT+omr or LJT+YFOSC), which are shown at the bottom of the plot.}
     \label{spec_cor}
\end{figure*}

\begin{figure*}
\centering
\includegraphics[scale=0.6]{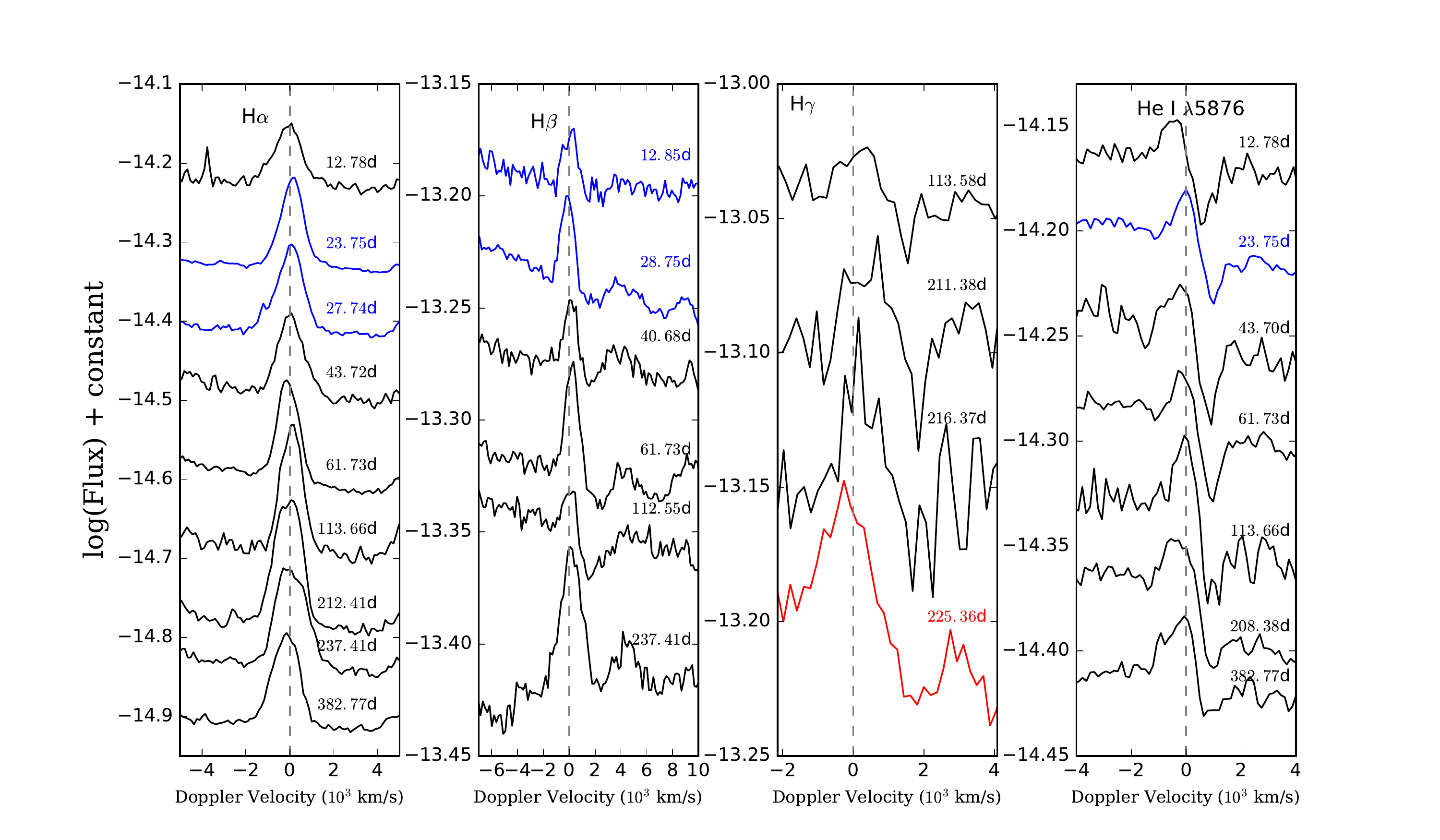}
\caption{The H$_{\alpha}$, H$_{\beta}$, H$_{\gamma}$, and He~{\sc i} $\lambda$5876 lines of MAXI J1820+070 in velocity space. The colors represent the spectra obtained with the instruments marked in Figure~\ref{spec_cor}.}
\label{lines}   
\end{figure*}

\begin{figure}
\centering
\includegraphics[scale=0.36]{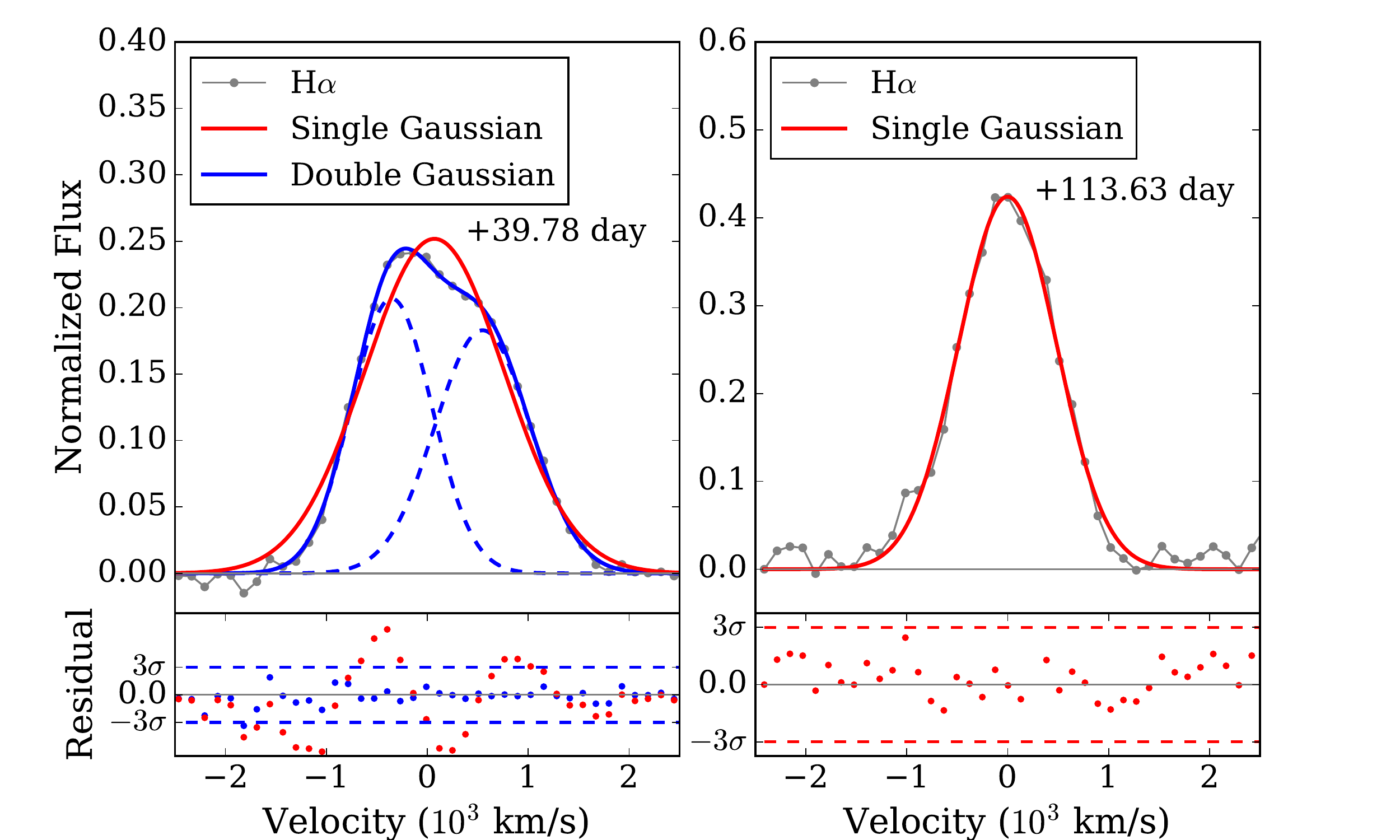}
\caption{The H$_\alpha$ line profile in two spectra of MAXI J1820+070, displayed in velocity space. Left panel: the H$_\alpha$ line profile in the t = +39.78 day spectrum. The red solid curve represents the fit curve using a single Gaussian function, while the blue solid line is the double Gaussian fit with two sub-components labeled by blue dashed lines. Right panel: the same case as in the left panel but for the t = +113.63 day spectrum. The lower panels show the residuals of the observed line profiles relative to the best-fit profiles.}
\label{doubleG}     
\end{figure}

Figure~\ref{spec_cor} shows the spectral evolution of MAXI J1820+070. The spectra are characterized by blue continua superimposed with H$_{\alpha}$ $\lambda$6563, H$_{\beta}$ $\lambda$4861, H$_{\gamma}$ $\lambda4341$, He~{\sc i} $\lambda$5876, He~{\sc i} $\lambda$6678, He~{\sc ii} $\lambda$4686 lines and Bowen blend. The detailed evolution of the main spectral features, i.e., H$_{\alpha}$, H$_{\beta}$, H$_{\gamma}$, and He~{\sc i} $\lambda$5876 are shown in Figure~\ref{lines}. The profiles of these features all show some degrees of temporal evolution, with some having only a single Gaussian component while others showing asymmetric line profiles, or double-peaked emission, as shown in Figure \ref{doubleG}. Using the Bayesian information criteria (BIC), a measure of the relative quality for a fit in model-selection \citep{10.1214/aos/1176344136}, we find that double Gaussian model (BIC=-548.8) works better than the single Gaussian model (BIC=-445.8) for H$_{\alpha}$ line at some phases (i.e., t=+39.78 day). However, in other phases (i.e., t=+113.63 day), the amplitudes of two sub-components of double Gaussian model differ so greatly that the weaker component is negligible. In this case, the single Gaussian model is still a better fitting model for such profiles. The single/double-peaked profile transition of emission lines was previously reported in many BH X-ray binaries \citep{2009MNRAS.393.1608F, 2018MNRAS.481.2646M}. A double-peaked line profile should originate from a rotating flow, i.e., the geometrically thin accretion discs \citep{1986MNRAS.218P...7M}, while winds from the accretion disc, driven by radiation pressure, tend to produce single-peaked lines \citep{1996Natur.382..789M}. However, the asymmetric single-line profile in our spectra can be formed by double Gaussian components which blended due to lower spectral resolution. Adding to the complexity, an apparently double-peaked emission line can be also produced from an intrinsically single-peaked profile with a central absorption \citep{2009MNRAS.393.1608F}. Given the above uncertainties in applying the double gaussian fit, we thus adopted the single gaussian fit in the following analysis of the main spectral features.

\subsection{Balmer Series}

\begin{figure}
   \includegraphics[width=\columnwidth]{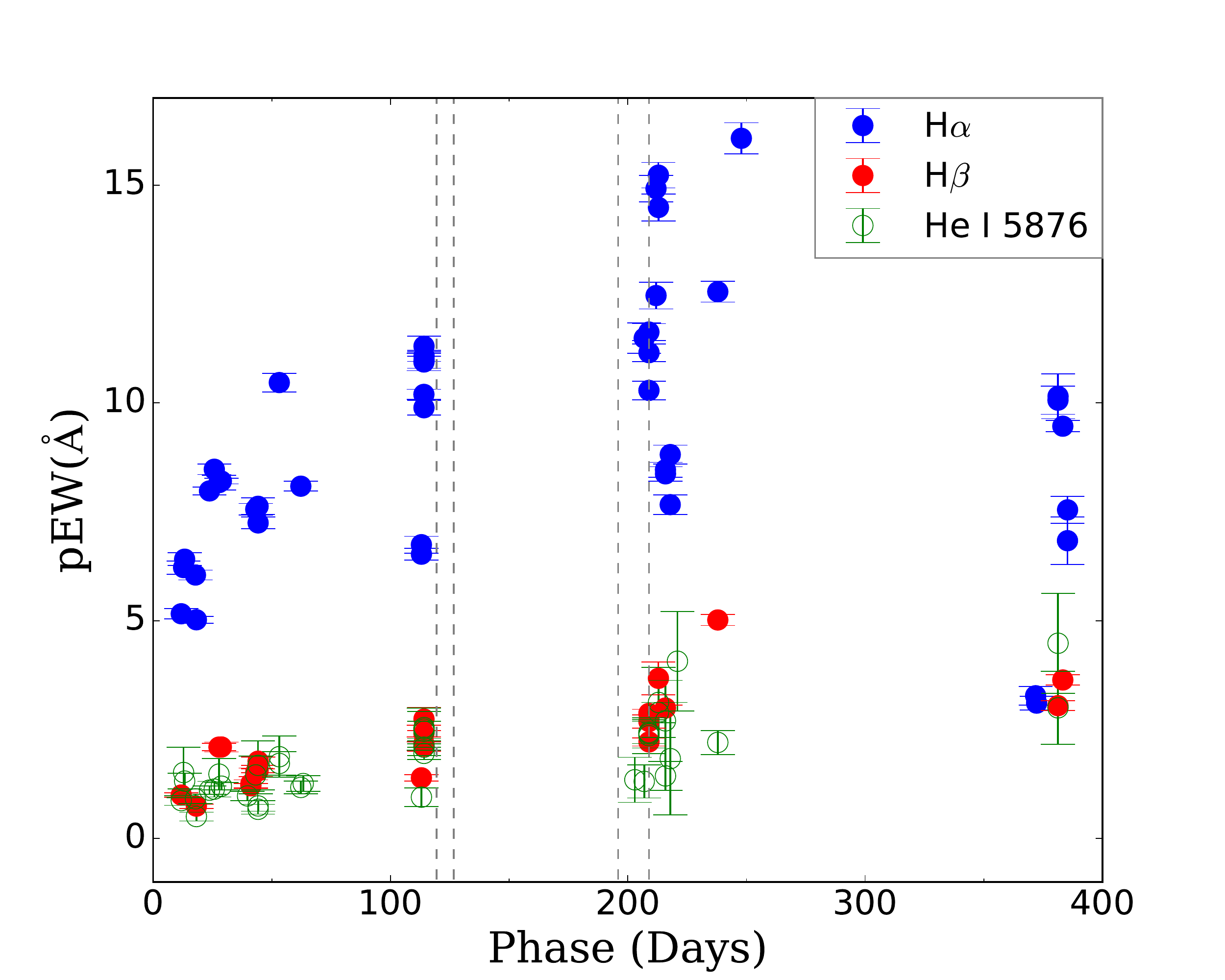}
       \caption{The pseudo equivalent width (pEW) evolution of the Balmer series and He~{\sc i} $\lambda$5876 in the spectra of MAXI J1820+070. All the lines were fit by single gaussian profiles. We use different colors to represent different spectral lines.}
     \label{figpew}
\end{figure}

\begin{figure}
\includegraphics[width=\columnwidth]{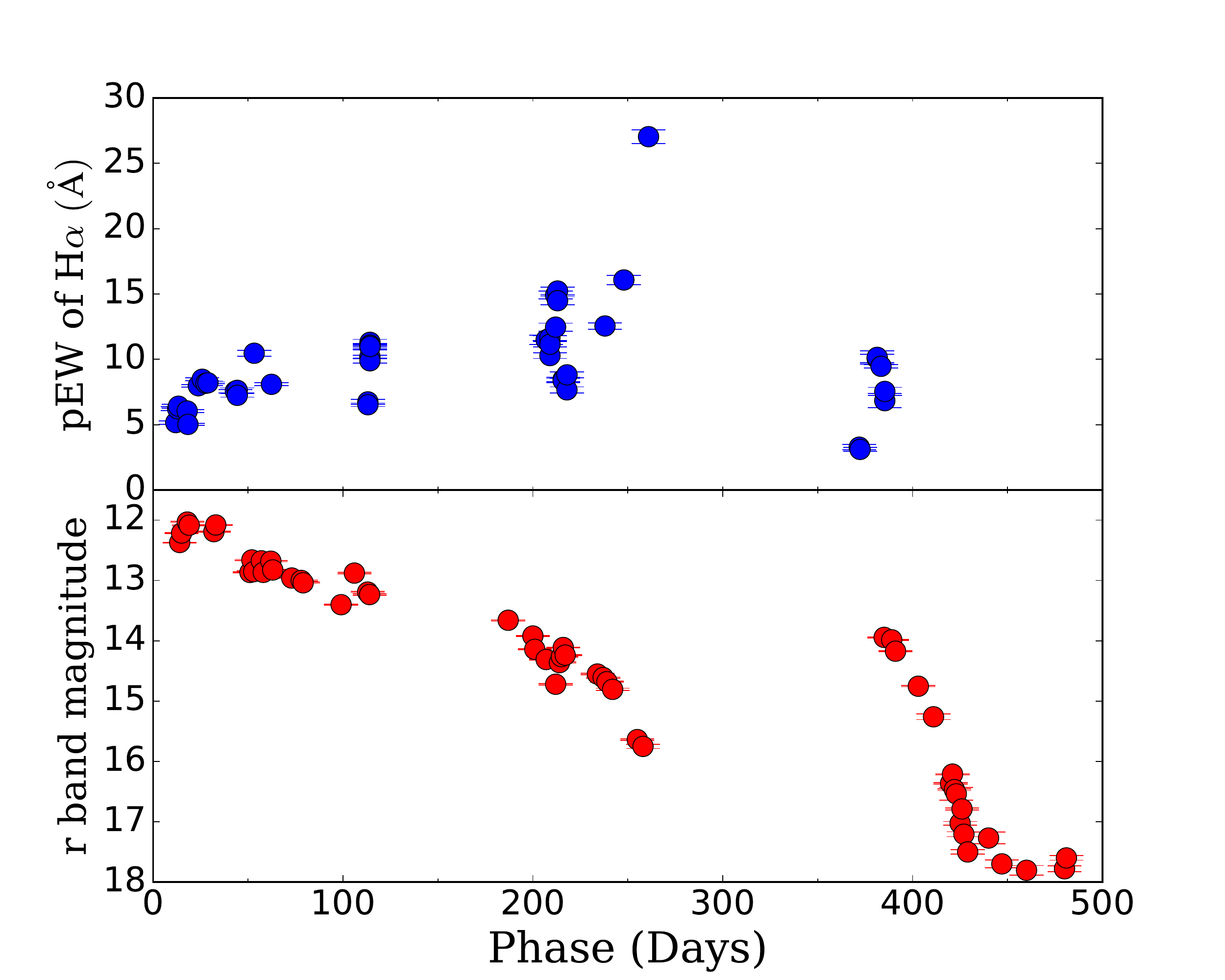}
\caption{The pseudo equivalent width evolution of H$_{\alpha}$ (top) and the $r$- band magnitude (bottom) measured for MAXI J1820+070. The time in days from the initial optical outburst are shown along the bottom axis.}
\label{havsr}     
\end{figure}

In Figure~\ref{figpew}, we present the temporal evolution of parameters measured from the Balmer emission lines. The local pseudo-continuum is determined via a linear fit to both sides of the feature; and repeated visual inspections are required for reducing the noise \citep{2016ApJ...826..211Z}. The pEW is calculated with the equation below:
\begin{equation}\label{pew}
\textup{pEW}=\int(F_{\textup{o}}-F_{\textup{c}})/F_{\textup{c}} d\lambda
\end{equation}
where $F_{\textup{o}}$ is the observed flux and $F_{\textup{c}}$ is the flux of the local pseudo-continuum.

Note that the H$_{\alpha}$ line shows a clear trend of strengthening within t$\sim$300 days from the initial detection of optical outburst, while its strength tends to decrease after t$\sim370$ days. The pEW of H$_{\beta}$ line holds a similar evolutionary trend as H$_{\alpha}$. Such an evolution pattern in Balmer lines can be explained with the change in optical depth as discussed in \cite{1980ApJ...235..939W} and \cite{2009MNRAS.393.1608F}. The spectrum of LMXB consists of a thermal continuum produced by inner hot, optically thick region of the disk, and emission lines formed in the outer cool region where the continuum is optically thin. During the rebrightening phase, the outer disc becomes hot as the X-ray irradiation gets stronger, and this makes the continuum become optically thick. The pEW of the emission lines depends on the relative size of the regions in the disk that are optically thin and thick in the continuum according to \cite{1980ApJ...235..939W}. Thus, the portion which is optically thick in the continuum increases during the rebrightening and this leads to decrease in pEW of optical lines from the outer region. The growth trend within $\sim$300 days can be also explained by this model. As the outer disk gets colder, an optically thin continuum forms at this stage and it leads to the increase in the pEWs of optical hydrogen emission lines.

To further examine the origin of the variation in H$_{\alpha}$ emission, we show in Figure~\ref{havsr} the correlation between the pEW of H$_{\alpha}$ and the $r$- band magnitude, which can characterize the strength of the continuum. The trend indicates that the pEW of H$_{\alpha}$ emission increases when the system becomes faint in optical, and vice versa. It is consistent with the optical depth explanation of the evolution pattern in Balmer lines.

\subsection{He~{\sc  i} $\lambda$5876 Feature}

He~{\sc i} $\lambda$5876 is found to show persistent P-Cygni profiles in the spectra of MAXI J1820+070 taken during the X-ray low/hard state, as similarly reported by \cite{2019ApJ...879L...4M}. Similar P-Cygni profile was also seen in V404 Cygni and other BH LMXBs \citep{2018MNRAS.481.2646M,2019MNRAS.488.1356C}, where the continuous presence of P-Cygni profiles were proposed to result from an expanding outflow formed during the outburst. One can notice that there is a prominent absorption feature on the red side of He~{\sc i} $\lambda$5876, which is likely due to the Galactic Na~{\sc i}~D absorption doublet. The evolution of He~{\sc i} $\lambda$5876 emission line parameters is plotted in Figure~\ref{figpew}, where one can see that the pEW follows a similar evolution trend as seen in H$_{\beta}$.

\section{Discussion}\label{sec:dis}
\subsection{SED and Temperature}
\begin{figure*}
\centering
\includegraphics[width=18cm]{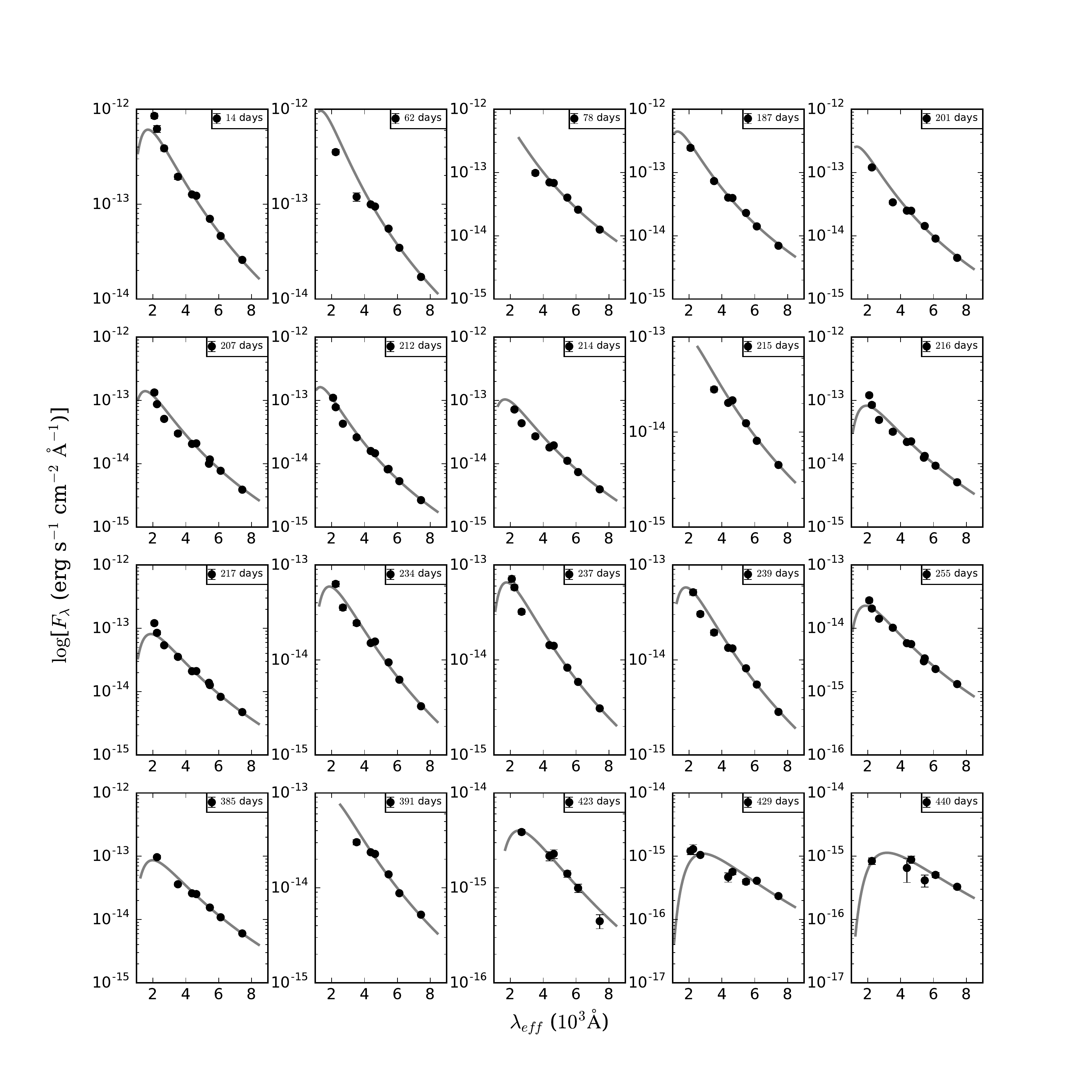}
\caption{Fits to the SEDs of MAXI J1820+070 using a single blackbody model. Data points are presented in filled circles and the grey lines show the best-fit models.}
\label{sed}     
\end{figure*}

\begin{figure}
          \centering
     \includegraphics[width=9cm]{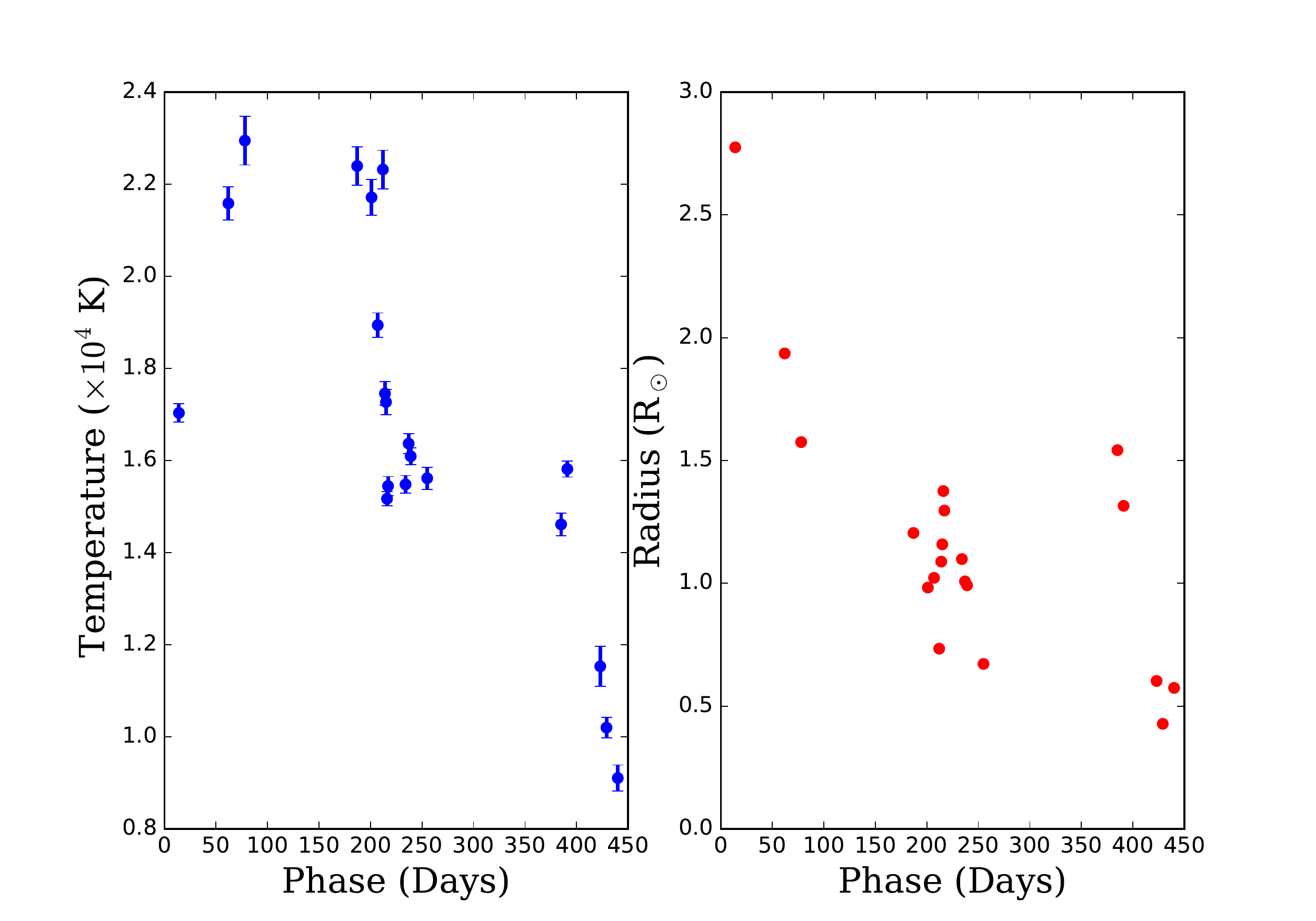}
\caption{Evolution of temperature (left) and radius (right) inferred for MAXI J1820+070 from the blackbody fits. The blue and red dots are the best-fit values of temperature and radius, respectively. }
\label{tr}
\end{figure}

To examine the radiation characteristics of MAXI J1820+070, we constructed the spectral energy distribution (SED) with the \emph{Swift}/UVOT, ground-based $B$, $g$, $V$, $r$, and $i$-band photometry to derive the temperature and radius of the photosphere, as shown in Figure~\ref{sed}. During the outburst phases, the optical and UV emissions from the companion star are negligible compared to those from the accretion disc (and possibly jets). The corresponding SEDs, corrected for the galactic extinction, is shown in Figure~\ref{sed}. It can be seen that in general the SEDs can not be well described by a single blackbody model, with the deviations mostly coming from the UV bands. This suggests that the emission regions of the optical and UV bands have different temperatures, i.e., a multi-temperature disc. However, fitting to the optical and UV bands separately would not provide useful constraints on different radiation zones because of limited data available for analysis. Contributions from the jets may also play a role, though they usually peak in the near-infrared bands \citep[e.g.,][]{2002ApJ...573L..35C,2006MNRAS.371.1334R}. 

\begin{figure}
\centering
\includegraphics[width=\columnwidth]{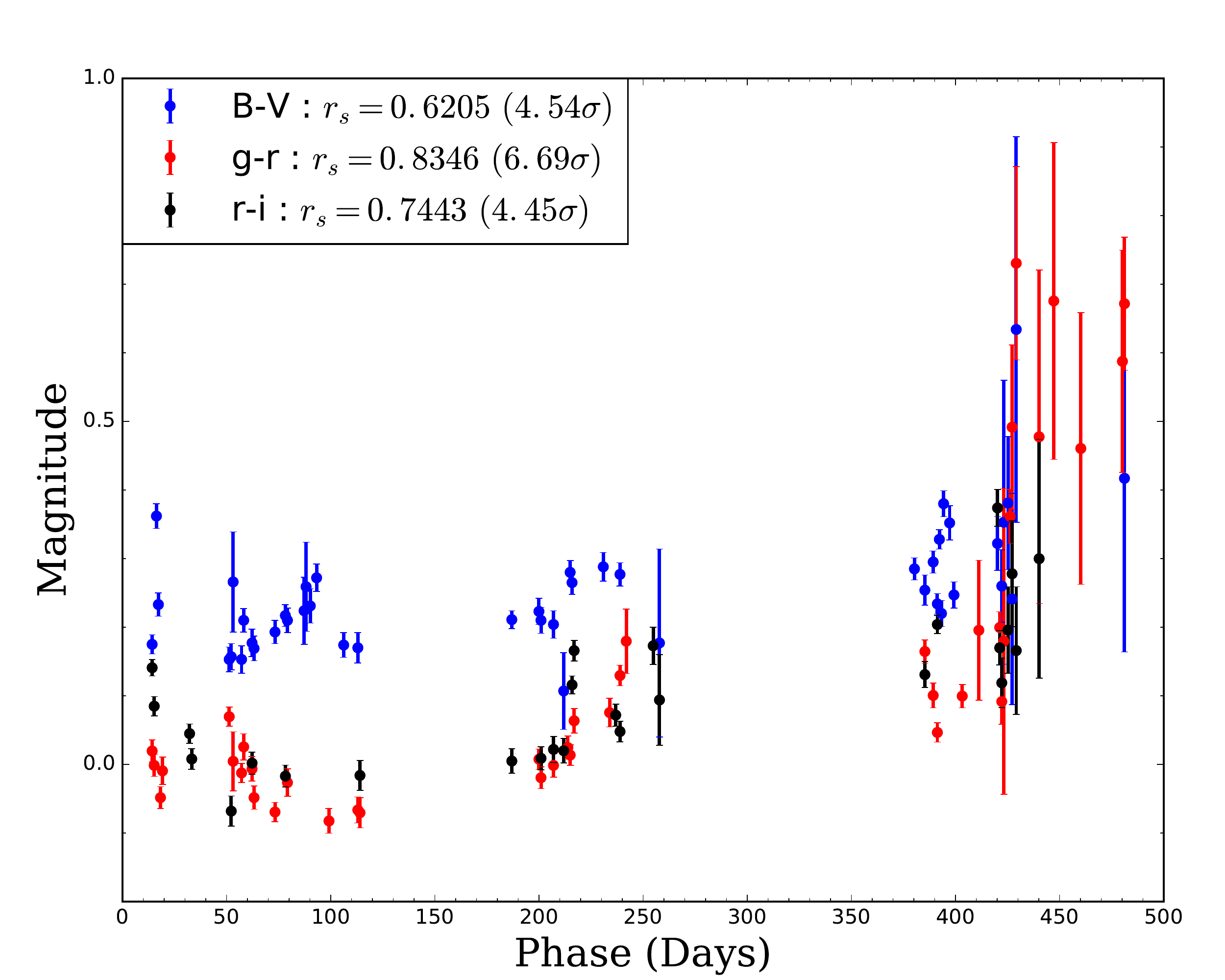}
\caption{The $B - V$, $g - r$, and $r - i$ color evolution of MAXI J1820+070. $r_s$ in the inset window represents the coefficient of the Spearman's correlation between the color evolution and the phase, while the number in the bracket represents the significance level.}
\label{bv_color}     
\end{figure}

Although the accretion disk may have multi-temperature components, the observed SED can still be reasonably fit by a simple blackbody model \citep{2018ApJ...867L...9T}, as shown in Figure~\ref{tr}. One can see that the blackbody temperature inferred from the multicolor photometry shows a rather complicated evolution, with an initial rapid increase from $\sim$17,000 K to 23,000 K, followed by a sudden drop to 15,000K at t $\sim$250 days. After a possible plateau evolution during t $\sim$100 - 200 days, the temperature then suffered another dramatic decline, i.e. from 15,000 K at t $\sim$250 days to 8,000 K at t $\sim$450 days. During the period from t$\sim$15 to $\sim$80 days after the initial outburst, MAXI J1820+070 appeared to become gradually hotter as its photosphere receded, which is consistent with the results from \cite{2018ApJ...867L...9T}. From t$\sim$180 to $\sim$260 days, the optical and UV emission region cooled down. This is consistent with the Balmer series and He~{\sc i} emission lines becoming stronger during similar phases (see Section \ref{sec:spectra}). After t$\sim$380 days, MAXI J1820+070 experienced a non-adiabatic evolution process, with the radiative radius becoming smaller while the temperature of the accretion disc keeping decreasing. Such a temperature decrease is consistent with the redwards evolution seen in the color curve (see Section \ref{sec:lc} and Figure \ref{bv_color}).


The $B-V$, $g-r$, and $r-i$ color curves of MAXI J1820+070, and the corresponding Spearman's rank correlation coefficient with phases are shown in Figure~\ref{bv_color}. The positive correlations indicate the post-outburst color tends to become progressively redder with a statistical significance of $\sim4.5\sigma$. This means the overall temperature of this system gradually decreased with time. However, the real physical process may be more complicated. There are some factors that cause the deviation from single blackbody, such as temperature difference in different regions of accretion disk and changes in optical depth.

\subsection{Optical and X-ray Relation}
\begin{figure}
\centering
\includegraphics[width=\columnwidth]{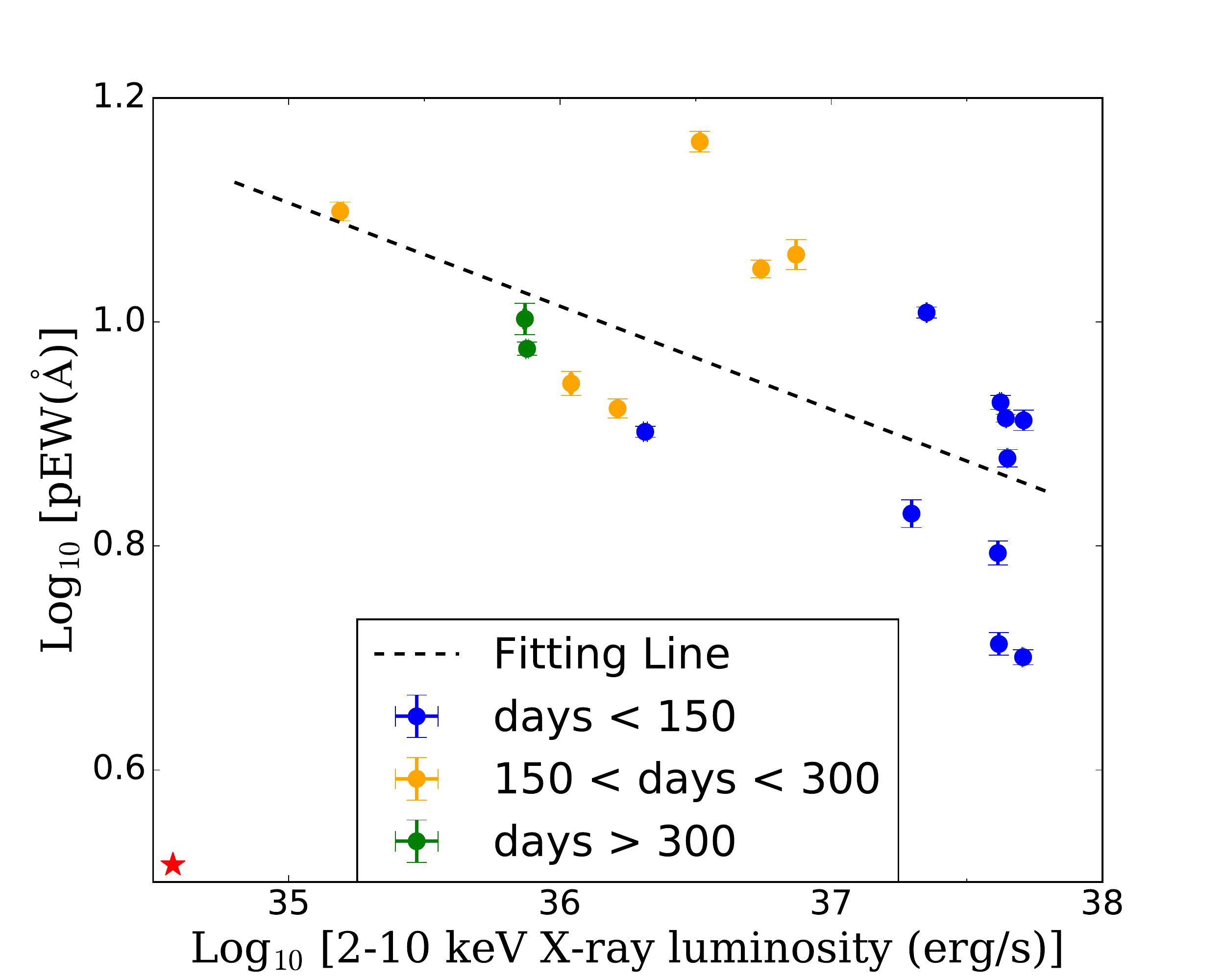}
\caption{Relation between the H$_{\alpha}$ pEW measured with the spectra presented in this paper and the X-ray luminosity inferred from the \emph{Swift} observations (2-10 keV). All the data are included in our fitting except the red star at the bottom left corner. Different colors represent different phases.}
\label{pew_xray}     
\end{figure}

\begin{figure*}
\centering
\includegraphics[scale=0.65]{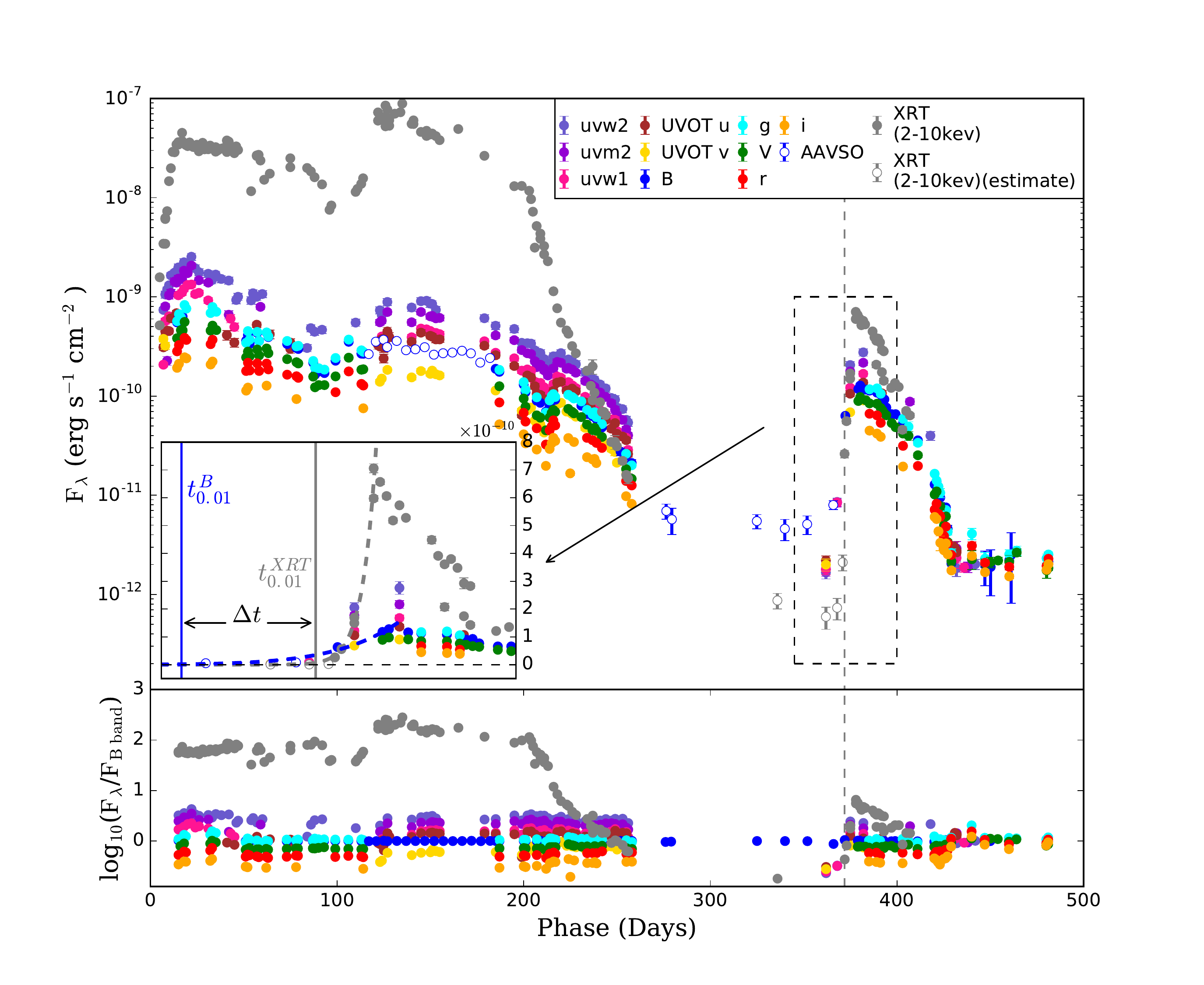}
\caption{The upper panel shows the optical, UV and X-ray light curves of MAXI J1820+070, which are all converted into flux, while the lower panel shows the logarithmic ratio of flux in each band to the flux in B band. Different colors represent different bands, including XRT (2-10kev), $uvw2$, $uvm2$, $uvw1$, UVOT $u$, UVOT $v$, $B$, $g$, $V$, $r$, and $i$ bands, as shown in the legend. The phase is defined in the same way as in Figure~\ref{long_lc}. The epoch of data represented by a red star in Figure~\ref{pew_xray} is marked by grey dashed line and the specific light curve around this time (345 - 400 days) is shown in the inset. The power-law fits to the rising light curves for XRT (2-10kev) and $B$ bands are shown in dashed lines with corresponding colors. The values of $t_{0.01}$and the time lag ($\Delta t$) measured for the rebrightening between $B$ band and XRT (2-10kev) , are also shown.}
\label{lc_flux}     
\end{figure*}

As mentioned in Section \ref{sec:intro}, MAXI J1820+070 is very bright in both optical and X-ray bands during the outburst. This gives us an opportunity to study the correlation between emission features in optical and the X-ray luminosity and put constraints on the physics of BH accretion.

In Figure~\ref{pew_xray}, we plot the pEW of H$_{\alpha}$ emission as a function of X-ray luminosity inferred for MAXI J1820+070 at comparable phases. We adopted the average value for the pEW of H$_{\alpha}$ line and X-ray luminosity when multiple observations are available on the same day. For MAXI J1820+070, most of these observations were obtained in the low/hard states. An anticorrelation can be seen between X-ray luminosity and the corresponding pEW of H$_{\alpha}$ with a confidence level of  $99.1\% (2.64\sigma)$. Based on a sample of six BH X-ray binaries, \cite{2009MNRAS.393.1608F} reported an anticorrelation between pEW of H$ _{\alpha}$ emission line and X-ray luminosity, which is $\textup{pEW} \propto L_x^{-0.18\pm0.06}$. Applying a power-law model to fit the observations of MAXI J1820+070, we get
\begin{equation}\label{ew-xray}
\textup{pEW}=10^{4.3\pm1.1}L_x^{-0.092\pm0.030}
\end{equation}
Note that this fit excludes the data point taken on t$\sim$+372 days. Inspecting the X-ray and optical/UV light curves shown in Figure~\ref{lc_flux}, one can see that at t$\sim$372 days, the optical and X-ray emission of MAXI J18020+070 was still in a rapid rising phase, and the BH binary system is likely in a transition state. This indicates that the anticorrelation between X-ray luminosity and H$_\alpha$ emission holds in the outburst phase but not in the transition phase. Thus the radiation mechanisms of the lines and disks during the transition phase are not consistent with that of the post-outburst phase.

\begin{figure*}
\centering
\includegraphics[scale=0.4]{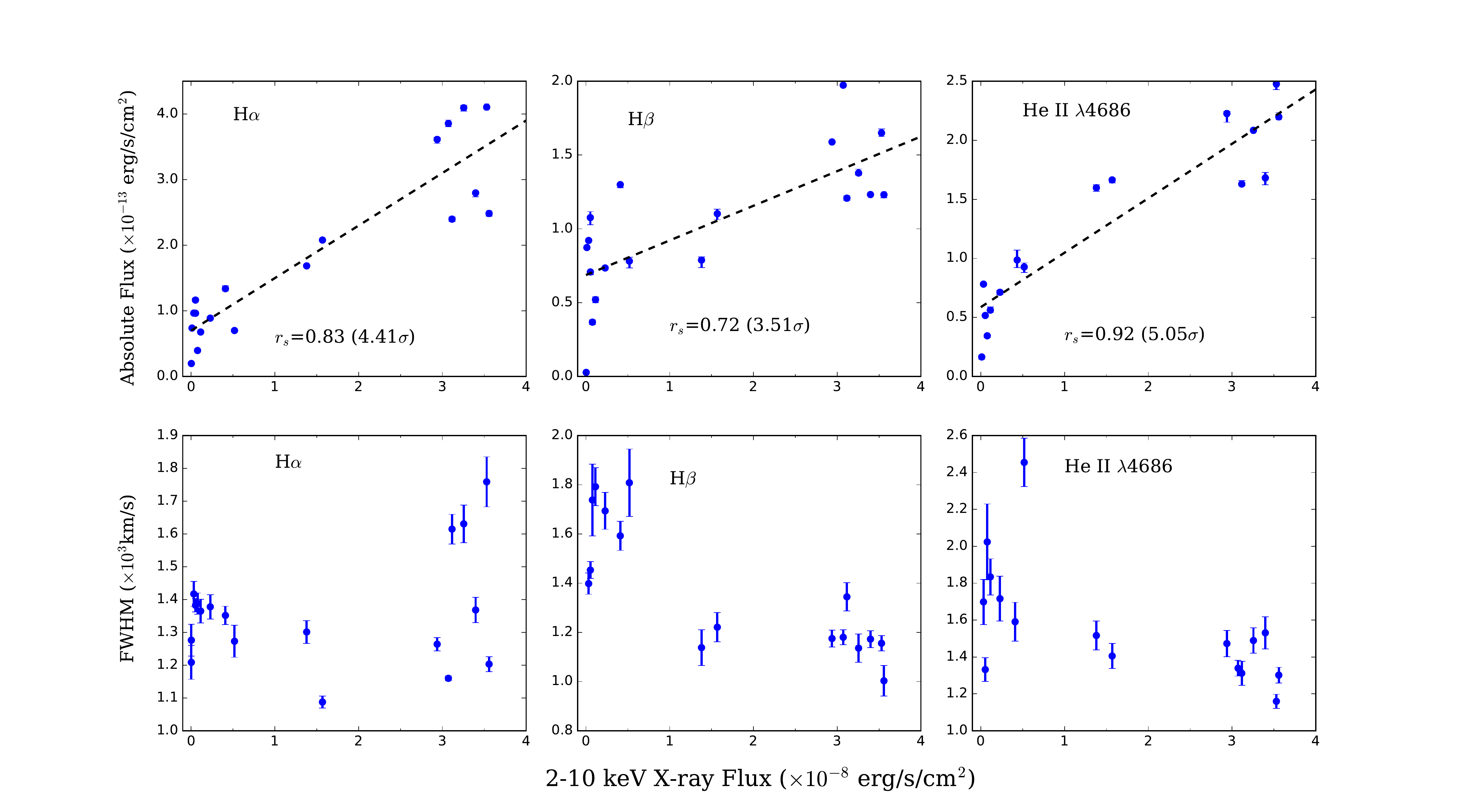}
\caption{Top panel: The relation between 2-10 keV X-ray flux and the absolute flux of H$_\alpha$, H$_\beta$ and He~{\sc ii} $\lambda$4686. The black dashed lines in above three panels represent the best linear fit to the data. The results of Spearman's rank correlation test was labeled in each panel. Bottom panel: The relation between 2-10 keV X-ray flux and the FWHM of H$_\alpha$, H$_\beta$ and He~{\sc ii} $\lambda$4686 lines.}
\label{xray_abs}     
\end{figure*}

As the measured pEW suffered from the effect of continuum flux, we further examine the relations between the absolute flux of lines such as H$_{\alpha}$, H$_{\beta}$, He~{\sc ii} $\lambda$4686 and 2-10 keV X-ray flux, as shown in Figure~\ref{xray_abs}. We find that the absolute fluxes of these lines show positive correlations with the X-ray radiation. We did not examine the correlation for He~{\sc i} $\lambda$5876 due to its possible blending with the Na I doublet. In comparison, the He~{\sc ii} $\lambda$4686 line seems to show the strongest correlation with the X-ray flux, because He~{\sc ii} has a higher ionization energy (i.e., 54.4eV) and can only be excited by X-ray photons \citep{2019ApJ...887..218L}.

Because the FWHM of H$_{\beta}$ and He~{\sc ii} $\lambda$4686 lines can be used to determine the outer edge of the accretion disk, we further examined their correlations with X-ray emission in lower-panel of Figure~\ref{xray_abs}. We found that with the increase of X-ray flux, both of these two lines tend to get weaker and then stabilize at 1200 km s$ ^\textup{-1}$ and 1400 km s$ ^\textup{-1}$, respectively. Based on the fact that the gas motion in accretion disk can be approximated to follow the Kepler motion, its velocity can be determined by $v=\sqrt{3/(8\ln2)}$~FWHM according to the analysis by \cite{netzer_2013}. The corresponding radius of the gas motion in the MAXI J1820+070 system can be estimated as $1.71\pm0.13 R_{\sun}$ from H$_{\beta}$ and $1.25\pm0.11 R_{\sun}$ from He~{\sc ii} $\lambda$4686 line, respectively. We find that both of these two values are less than the typical radius of the accretion disk surrounding the stellar-mass BH \citep{2001NewAR..45..449L}, which equals to the 90\% of effective Roche radius $R_L=2.46R_{\sun}$, identified by the relevant parameters obtained by  \cite{2019ApJ...882L..21T}. The larger radius inferred from H$_\beta$ relative to He~{\sc ii} $\lambda$4686 indicates that the formation region of the former line can extend to the outer side of the accretion disc. This is due to the fact that He~{\sc ii} $\lambda$4686 has a higher excitation energy than H$_\beta$. On the outer part of the accretion disk with lower temperature, only H$_\beta$ can be excited. Note, however, that H$_\alpha$ did not show such a similar relationship with X-ray flux. This suggests that the formation region of H$_\alpha$ line was not confined to the accretion disk, but also included the outflow, as indicated by the P-cygni profile of H$_\alpha$ from \cite{2019ApJ...879L...4M}.

The thermal-viscous DIM describes that, in quiescence, the accretion disc is filled up with matter, until at some radius, its temperature reaches the critical value that ionizes the hydrogen to trigger an outburst \citep{2001NewAR..45..449L, 2019AN....340..278R}. This outburst can be triggered within the inner part of the accretion disc, and then propagate outwards, which is called as ``inside-out" outburst \citep{2001NewAR..45..449L}. However, in other LMXBs, the triggering process may start at the outer disc and then propagates inwards to the inner region, which is called as ``outside-in" outburst \citep{1998MNRAS.300.1035S}. Optical emission is usually produced by the outer disc, while the X-ray emission can be created on the viscous timescale when the inner disc is filled in. Thus detecting the time lag between different bands helps determine where the instability is triggered in the accretion disk.

Based on the method introduced in \cite{2018ApJ...867L...9T}, we use a power-law function to fit the rising light curves,
\begin{equation}\label{powerlaw}
\hat{F}_{\lambda}(t) = A\left (\frac{t-t_{0,\lambda}}{1~\textup{day}}\right)^{B}
\end{equation}
where $\hat{F}_{\lambda}$ is normalized flux for each band, $t_{0,\lambda}$ is the zero point assumed for the reference epoch and $t$ is the time of photometry.  We focus on the time $t_{0.01}$, representing the epoch when the light curve reaches 1\% of the peak flux. \cite{2018ApJ...867L...9T} found that during the initial outburst, the rising in optical preceded that in the X-ray by $\Delta t=t_{0.01}^{BAT}-t_{0.01}^{V}=7.20 \pm 0.97$ days, and here we turn to study the time-lag during the rebrightening period from t $\sim$ 370 to 410 days. The power-law fits to the rebrightening evolution in $B$ band and \emph{Swift} XRT 2-10kev band are plotted in the inset of Figure~\ref{lc_flux}. Based on the fitting parameters and bootstrap-resampling technique of estimating errors, we get $t_{0.01}^{XRT} = 58.92 \pm 0.15$ days for \emph{Swift} XRT (2-10kev) and $t_{0.01}^{B} = 38.12 \pm 2.70$ days for the $B$ band. Then we can determine a time lag between optical and X-ray outbursts as $\Delta t=t_{0.01}^{XRT}-t_{0.01}^{B}=20.80 \pm 2.85$ ($\Delta t \pm2\sigma$) days. This means that during the rebrightening process, the start of the rising in optical preceded that in the X-ray, which is consistent with the estimate for the first outburst from \cite{2018ApJ...867L...9T}. The time-lag allows constraints on the radius $R(B)$ for the disk where the outburst is triggered and propagates simultaneously both inward and outward until reaching the inner region and producing the X-rays. We use the scaling relation in \cite{2016ApJ...818L...5B} and the same parameter setup adopted in \cite{2018ApJ...867L...9T} to find the radius as $R(B)=[0.025-0.079]R_{\sun}$ ($2\sigma$), which for the initial outburst is $R(V) = [0.01-0.04] R_{\sun}$, approval of the DIM and irradiation models for BH LMXB.

For MAXI J1820+070, one interesting evolution occurred at t$\sim$200 to 300 days, when the emission is found to show much larger decrease in X-ray than in optical and UV bands (see also Figure~\ref{lc_flux}). This observation fact suggests that the effects of irradiation on the accretion disk are diminishing during this phase, and the viscous energy of the accretion disk may contribute to maintain a lower rate of decline. Moreover, in MAXI J1820+070, an intensity jump appears in optical and UV bands at t$\sim$210 days. \cite{1993A&A...279L..13A} reported that some LMXBs exhibited intensity jumps during the decay of the initial outburst, and they interpreted this phenomenon as an instantaneous response of the companion to the heating of X-rays and a response of the disk to the extra mass flow. The amplitude of this intensity jump tends to get stronger in redder bands (and vice versa). Note, however, this intensity jump did not occur in the X-ray band. This can be explained by a temporary increase in accretion rate of the outer disk, while this change will not affect the inner disk due to a longer viscous time scale.

\section{Conclusions}\label{sec:con}
In this work, we present optical, UV and X-ray observations and studies for the BH binary system MAXI J1820+070. Several outbursts/rebrightenings were recorded during our monitoring campaign spanning from its initial optical outburst to $\sim$550 days after. The main results from our analysis of the data are summarized as follows. 

\begin{enumerate}

\item The spectra of MAXI J1820+070 are characterized by blue continuum and the superimposed Balmer series, helium emission lines, and Bowen blend features, similar to those seen in several other well-known BH LMXBs like V404 Cygni \citep{2018MNRAS.481.2646M}. The H$_{\alpha}$ emission becomes progressively strong within $\sim$300 days from the first detection of outburst, followed by a subsequent weakening during 350$\sim$400 days. A similar trend can be also found in the H$_{\beta}$ feature. The He~{\sc i} $\lambda$5876  line is also found to become progressively stronger during the phase from the first detection to about 400 days after that. Such an evolution trend can be explained by the change in optical depth of continuum, as a result of  the temperature change of the outer disc during the outburst.

\item Based on extensive optical spectroscopy and the X-ray observations of MAXI J18020+070, we analyzed the relationship between the X-ray luminosity and the pEW of the H$_{\alpha}$ emission, and we found an anticorrelation as $\textup{pEW}_{H_{\alpha}}=10^{4.3\pm1.1}L_x^{-0.092\pm0.030}$, confirming previous results.

\item After removing the effect of continuum, the absolute flux of H$_\alpha$, H$_{\beta}$ and He~{\sc ii} $\lambda$4686  showed positive correlations with the X-ray flux obtained at similar phases. This confirms that the high-energy irradiation can be regarded the pumping source for the emission of MAXI1820+070. In addition, the FWHM of H$_{\beta}$ and He~{\sc ii} $\lambda$4686 is found to become smaller with the increase of X-ray flux but they tend to stabilize at line forming region of $1.25-1.71R_{\sun}$, consistent with the typical radius of accretion disk around a stellar-size black hole.

\item The time lag found for the change in emission between optical and X-ray bands is $\Delta t=t_{0.01}^{XRT}-t_{0.01}^{B}=20.80 \pm 2.85$ ($\Delta t \pm2\sigma$) days, which means that during the rebrightening process, the start of the rising in optical preceded that in the X-ray. 

\item At t$\sim$200-300 days, the X-ray flux is found to show a sudden drop, while the flux variation in optical/UV flux is much less significant. This discrepancy suggests that the viscous energy of the accretion disk can contribute significantly to the optical/UV flux when irradiation diminished.

\end{enumerate}

\section*{Acknowledgements}

We thank the anonymous referee for suggestive comments to help improve this work. We acknowledge the support of the staff of the Lijiang 2.4 m and Xinglong 2.16 m telescopes. Funding for the LJT has been provided by Chinese Academy of Sciences and the People's Government of Yunnan Province. The LJT is jointly operated and administrated by Yunnan Observatories and the Center for Astronomical Mega-Science, CAS. We acknowledge the support of the staff of the Xinglong 80cm telescope. It was partially supported by the Open Project Program of the Key Laboratory of Optical Astronomy, National Astronomical Observatories, Chinese Academy of Sciences. This work is supported by the National Natural Science Foundation of China (NSFC grants 12033003, 11761141001 and 11633002), the National Program on Key Research and Development Project (grant no. 2016YFA0400803). X.W. is also supported by the Scholar Program of Beijing Academy of Science and Technology (BS2020002). J.W. acknowledges the support of the National Natural Science Foundation of China (grant No. U1938105) and the President Fund of Xiamen University (No. 20720190051). Lingzhi Wang is sponsored (in part) by the Chinese Academy of Sciences (CAS), through a grant to the CAS South America Center for Astronomy (CASSACA) in Santiago, Chile. CYW is supported by the National Natural Science Foundation of China (NSFC grants 12003013). This research has made use of data provided by the Yaoan High Precision Telescope. We thank the staff of AZT for their observations and allowance of the use of the data. We acknowledge with thanks the variable star observations from the AAVSO International Database contributed by observers worldwide and used in this research.

\textit{software}: ZrutyPhot (Mo et al. in prep.), IRAF \citep{1986SPIE..627..733T,1993ASPC...52..173T}, SExtractor \citep{1996A&AS..117..393B}    

\section*{Data availability}
The data underlying this article are available in the article. Our photometric data has been attached in Table~\ref{standard},~\ref{table_lc} and \ref{table_lc_uv} in the appendix. The log table of spectral data is also shown in Table~\ref{table_log}.


\bibliographystyle{mnras}
\bibliography{mnras} 



\appendix

\section{PHOTOMETRIC AND SPECTROSCOPIC DATA}
\begin{table*}
\centering
\caption{Photometric Standards in the MAXI J1820+070 Field}
\begin{tabular}{ccccccccccc} 
\hline
Num. &$\alpha$(J2000) &
$\delta$(J2000) & \textit{B} (mag) & \textit{V} (mag) &\textit{g} (mag)& \textit{r} (mag)&\textit{i} (mag)\\	
\hline
	        
1 & $18^\mathrm{h}20^\mathrm{m}25^\mathrm{s}.4256$ & $7\degr09\arcmin21\arcsec.6576$ & 14.524(068) & 13.828(032) & 14.081(002) & 13.647(001) & 13.437(001)\\ 
2 & $18^\mathrm{h}20^\mathrm{m}32^\mathrm{s}.2418$& $7\degr15\arcmin02\arcsec.0196$ & 14.534(091) & 13.862(037) & 14.111(006) & 13.701(005) & 13.542(003)\\ 
3 & $18^\mathrm{h}20^\mathrm{m}40^\mathrm{s}.1472$ & $7\degr10\arcmin57\arcsec.8100$ & 15.361(090) & 13.808(033) & 14.419(001) & 13.667(030) & 12.982(...)\\ 
4 & $18^\mathrm{h}20^\mathrm{m}26^\mathrm{s}.4202$ & $7\degr10\arcmin11\arcsec.7948$ & 13.976(095) & 12.796(046) & ... & ... & ...\\ 
5 & $18^\mathrm{h}20^\mathrm{m}27^\mathrm{s}.5782$ & $7\degr11\arcmin40\arcsec.8696$ & 14.697(101) & 13.622(039) & ... & ... & ...\\ 
6 & $18^\mathrm{h}20^\mathrm{m}25^\mathrm{s}.6668$ & $7\degr15\arcmin51\arcsec.2424$ & 14.037(101) & 12.805(044)  & ... & ... & ...\\ 
7 & $18^\mathrm{h}20^\mathrm{m}40^\mathrm{s}.0294$ & $7\degr12\arcmin01\arcsec.9044$ & 14.778(091) & 13.322(036) & ... & ... & ...\\ 

8 & $18^\mathrm{h}20^\mathrm{m}31^\mathrm{s}.2221$ & $7\degr06\arcmin48\arcsec.8412$ & 13.792(077) & 12.285(030) & ... & ... & ...\\ 
9 & $18^\mathrm{h}20^\mathrm{m}07^\mathrm{s}.1098$ & $7\degr13\arcmin28\arcsec.6068$ & 14.909(096) & 14.286(045) & ... & ... & ...\\ 
10 &  $18^\mathrm{h}20^\mathrm{m}10^\mathrm{s}.1870$  &  $7\degr13\arcmin50\arcsec.7982 $ & ... & ... &14.759(002) & 14.189(002) & 13.960(001) \\ 
11 &  $18^\mathrm{h}20^\mathrm{m}32^\mathrm{s}.1594$  &  $7\degr07\arcmin04\arcsec.6632$  & ... & ... & 14.760(002) & 14.201(001) & 13.975(000) \\ 
12 &  $18^\mathrm{h}20^\mathrm{m}18^\mathrm{s}.5156$  &  $7\degr10\arcmin32\arcsec.7939$& ... & ... & 14.851(003) & 14.051(001) & 13.687(005) \\ 
\hline
\multicolumn{5}{l}{Note: Uncertainties, in units of 0.001 mag, are $1\sigma$.}\\
\end{tabular}
\label{standard} 
\end{table*}

\begin{table*}
\centering
\caption{Ground-based Optical Photometry of  MAXI J1820+070 \label{gphoto} }
\begin{tabular}{ccccccccccccc} 
\hline
MJD &Phase$^a$ &\textit{B} (mag)& \textit{g} (mag)& \textit{V} (mag) &\textit{r} (mag) & \textit{R} (mag)&  \textit{i} (mag) & data source \\	
\hline
        58198.3 & 14.2 & 12.573(005) & 12.390(010) & 12.398(009) & 12.371(006) & ... & 12.230(006) & TNT \\ 
        58199.3 & 15.2 & 12.430(008) & 12.212(008) & ... & 12.214(008) & ... & 12.129(006) & TNT \\ 
        58200.4 & 16.3 & 12.558(012) & 12.294(014) & 12.196(006) & ... & ... & ... & TNT \\ 
        58201.3 & 17.2 & 12.433(007) & ... & 12.200(010) & ... & ... & ... & TNT \\ 
        58202.3 & 18.2 & ... & 11.983(007) & ... & 12.032(009) & ... & 11.965(010) & TNT \\ 
        58202.4 & 18.3 & ... & ... & 11.987(007) & ... & ... & ... & TNT \\ 
        58203.3 & 19.2 & ... & 12.074(010) & ... & 12.084(010) & ... & 11.991(009) & TNT \\ 
        58216.3 & 32.2 & ... & 12.185(006) & 12.206(007) & 12.191(009) & ... & 12.146(005) & TNT \\ 
        58217.3 & 33.2 & ... & 12.034(006) & 12.081(005) & 12.080(006) & ... & 12.072(009) & TNT \\ 
        58219.4 & 35.3 & ... & 12.153(007) & 12.193(008) & ... & ... & ... & TNT \\ 
        58235.3 & 51.2 & 13.034(008) & 12.936(007) & 12.881(010) & 12.867(007) & ... & 12.805(008) & TNT \\ 
        58236.3 & 52.2 & 12.892(011) & 12.641(011) & 12.735(008) & 12.660(011) & ... & 12.728(011) & TNT \\ 
        58237.2 & 53.1 & 13.118(040) & 12.858(030) & 12.852(033) & 12.854(013) & ... & ... & TNT \\ 
        58241.2 & 57.1 & 12.836(012) & 12.662(007) & 12.683(008) & 12.675(007) & ... & ... & TNT \\ 
        58242.3 & 58.2 & 13.042(009) & 12.891(008) & 12.832(008) & 12.866(011) & ... & ... & TNT \\ 
        58246.3 & 62.2 & 12.834(011) & 12.674(009) & 12.657(009) & 12.681(009) & ... & 12.679(007) & TNT \\ 
        58247.2 & 63.1 & 12.945(011) & 12.777(010) & 12.776(007) & 12.826(007) & ... & ... & TNT \\ 
        58257.3 & 73.2 & 13.119(010) & 12.890(007) & 12.926(007) & 12.960(007) & ... & ... & TNT \\ 
        58262.3 & 78.2 & 13.220(009) & 13.025(008) & 13.003(007) & 13.000(009) & ... & 13.017(007) & TNT \\ 
        58263.3 & 79.2 & 13.244(010) & 13.012(011) & 13.034(008) & 13.039(009) & ... & ... & TNT \\ 
        58271.3 & 87.2 & 13.591(028) & 13.394(019) & 13.367(021) & ... & ... & ... & TNT \\ 
        58272.2 & 88.1 & 13.894(036) & 13.565(025) & 13.635(029) & ... & ... & ... & TNT \\ 
        58274.2 & 90.1 & 13.811(013) & 13.578(008) & 13.580(012) & ... & ... & ... & TNT \\ 
        58277.3 & 93.2 & 13.852(011) & 13.612(010) & 13.580(009) & ... & ... & ... & TNT \\ 
        58283.2 & 99.1 & 13.540(008) & 13.319(009) & 13.360(007) & 13.402(009) & ... & ... & TNT \\ 
        58290.3 & 106.2 & 13.063(007) & 12.855(007) & 12.889(011) & 12.878(010) & ... & ... & TNT \\ 
        58297.1 & 113.0 & 13.362(012) & 13.126(009) & 13.192(010) & 13.193(010) & ... & ... & TNT \\ 
        58298.1 & 114.0 & 13.366(013) & 13.163(010) & 13.181(011) & 13.234(012) & ... & 13.250(010) & TNT \\ 
        58300.5 & 116.4 & 13.374(010) & ... & ... & ... & ... & ... & AAVSO \\ 
        58304.3 & 120.2 & 13.061(011) & ... & ... & ... & ... & ... & AAVSO \\ 
        58308.4 & 124.3 & 13.010(065) & ... & ... & ... & ... & ... & AAVSO \\ 
        58310.4 & 126.3 & 13.197(045) & ... & ... & ... & ... & ... & AAVSO \\ 
        58315.5 & 131.4 & 13.042(007) & ... & ... & ... & ... & ... & AAVSO \\ 
        58320.5 & 136.4 & 13.276(009) & ... & ... & ... & ... & ... & AAVSO \\ 
        58325.5 & 141.4 & 13.253(009) & ... & ... & ... & ... & ... & AAVSO \\ 
        58330.5 & 146.4 & 13.200(014) & ... & ... & ... & ... & ... & AAVSO \\ 
        58335.4 & 151.3 & 13.388(028) & ... & ... & ... & ... & ... & AAVSO \\ 
        58340.4 & 156.3 & 13.352(026) & ... & ... & ... & ... & ... & AAVSO \\ 
        58345.3 & 161.2 & 13.335(013) & ... & ... & ... & ... & ... & AAVSO \\ 
        58350.5 & 166.4 & 13.289(010) & ... & ... & ... & ... & ... & AAVSO \\ 
        58355.3 & 171.2 & 13.355(017) & ... & ... & ... & ... & ... & AAVSO \\ 
        58360.3 & 176.2 & 13.588(014) & ... & ... & ... & ... & ... & AAVSO \\ 
        58365.5 & 181.4 & 13.473(012) & ... & ... & ... & ... & ... & AAVSO \\ 
        58369.0 & 184.9 & 13.745(008) & ... & ... & ... & ... & ... & TNT \\ 
        58371.0 & 186.9 & 13.820(006) & 13.618(007) & 13.609(007) & 13.660(010) & ... & 13.655(008) & TNT \\ 
        58384.0 & 199.9 & 14.142(008) & 13.928(007) & 13.919(011) & 13.921(008) & ... & ... & TNT \\ 
        58384.1 & 200.0 & ... & ... & ... & ... & ... & 13.907(010) & TNT \\ 
        58385.0 & 200.9 & 14.334(008) & 14.122(007) & 14.124(011) & 14.142(009) & ... & 14.133(008) & TNT \\ 
        58391.0 & 206.9 & 14.547(010) & 14.309(008) & 14.343(010) & 14.311(009) & ... & 14.289(010) & TNT \\ 
        58393.0 & 208.9 & 14.602(010) & ... & 14.400(006) & ... & ... & ... & TNT \\ 
        58395.9 & 211.8 & 14.826(037) & 14.698(013) & 14.719(019) & ... & ... & ... & TNT \\ 
        58396.0 & 211.9 & ... & ... & ... & 14.721(011) & ... & 14.701(007) & TNT \\ 
        58397.9 & 213.8 & 14.687(030) & 14.387(008) & 14.397(010) & 14.363(009) & ... & ... & TNT \\ 
        58398.0 & 213.9 & ... & ... & ... & ... & ... & 14.274(010) & TNT \\ 
        58399.0 & 214.9 & 14.560(010) & 14.276(006) & 14.280(007) & 14.263(009) & ... & ... & TNT \\ 
        58399.1 & 215.0 & ... & ... & ... & ... & ... & 14.129(010) & TNT \\ 
        58400.0 & 215.9 & 14.470(010) & 14.230(007) & 14.205(007) & 14.112(007) & ... & 13.996(006) & TNT \\ 
        58401.0 & 216.9 & 14.523(010) & 14.301(010) & 14.253(010) & 14.238(008) & ... & 14.072(007) & TNT \\ 
        58408.0 & 223.9 & ... & 14.246(007) & 14.179(007) & ... & ... & 14.000(007) & TNT \\ 
        58409.0 & 224.9 & ... & ... & ... & ... & ... & 14.890(054) & TNT \\ 
        58410.0 & 225.9 & ... & 14.302(008) & 14.254(009) & ... & ... & ... & TNT \\ 
        58411.0 & 226.9 & ... & 14.340(008) & 14.291(008) & ... & ... & 14.099(008) & TNT \\ 
\hline
\multicolumn{5}{l}{$^a$ Days relative to the initial optical outburst (UT March 06.58 2018 = MJD 58184.08).}\\
\multicolumn{5}{l}{Note: Uncertainties, in units of 0.001 mag, are $1\sigma$.}\\
\end{tabular}
\label{table_lc} 
\end{table*}

\begin{table*}
 \contcaption{A table continued from the previous one.}
 \label{tab:continued}
 \begin{tabular}{ccccccccc}
 \hline
MJD &Phase$^a$ &\textit{B} (mag)& \textit{g} (mag)& \textit{V} (mag) &\textit{r} (mag) & \textit{R} (mag)&  \textit{i} (mag) & data source \\	
  \hline
        58415.0 & 230.9 & 14.674(012) & 14.413(008) & 14.386(009) & ... & ... & ... & TNT \\ 
        58418.0 & 233.9 & 14.883(010) & 14.627(011) & 14.575(011) & 14.552(010) & ... & 14.482(010) & TNT \\ 
        58420.9 & 236.8 & 14.938(026) & 14.739(011) & 14.715(011) & 14.610(009) & ... & 14.538(007) & TNT \\ 
        58422.0 & 237.9 & ... & 14.685(009) & 14.655(009) & ... & ... & 14.526(011) & TNT \\ 
        58423.0 & 238.9 & 15.009(009) & 14.808(007) & 14.732(008) & 14.679(008) & ... & 14.631(007) & TNT \\ 
        58426.0 & 241.9 & 15.223(030) & 14.985(027) & 14.887(025) & 14.806(020) & ... & ... & TNT \\ 
        58436.0 & 251.9 & 15.833(095) & ... & ... & ... & ... & ... & TNT \\ 
        58438.9 & 254.8 & 15.916(029) & 15.735(014) & 15.699(015) & 15.639(013) & ... & 15.466(014) & TNT \\ 
        58441.9 & 257.8 & 16.119(090) & 16.024(056) & 15.942(047) & 15.752(034) & ... & 15.658(032) & TNT \\ 
        58523.5 & 339.4 & 17.776(264) & ... & ... & ... & ... & ... & AAVSO \\ 
        58535.4 & 351.3 & 17.661(230) & ... & ... & ... & ... & ... & AAVSO \\ 
        58549.4 & 365.3 & 17.177(108) & ... & ... & ... & ... & ... & AAVSO \\ 
        58556.4 & 372.3 & 14.930(009) & ... & ... & ... & ... & ... & TNT \\ 
        58563.4 & 379.3 & 14.254(011) & ... & 13.980(009) & ... & ... & ... & TNT \\ 
        58564.4 & 380.3 & 14.167(009) & ... & 13.882(007) & ... & ... & ... & TNT \\ 
        58567.4 & 383.3 & 14.269(007) & ... & 13.960(007) & ... & ... & ... & TNT \\ 
        58569.4 & 385.3 & 14.294(012) & 14.108(009) & 14.040(010) & 13.944(008) & ... & 13.813(011) & TNT \\ 
        58573.4 & 389.3 & 14.345(007) & 14.085(009) & 14.050(009) & 13.985(009) & ... & 13.888(009) & TNT \\ 
        58575.4 & 391.3 & 14.391(007) & 14.219(008) & 14.157(008) & 14.173(006) & ... & 13.969(007) & TNT \\ 
        58576.3 & 392.2 & 14.548(006) & ... & 14.220(008) & ... & ... & ... & TNT \\ 
        58577.4 & 393.3 & 14.516(010) & ... & 14.296(009) & ... & ... & ... & TNT \\ 
        58578.4 & 394.3 & 14.720(011) & ... & 14.340(008) & ... & ... & ... & TNT \\ 
        58581.3 & 397.2 & 14.889(012) & ... & 14.537(013) & ... & ... & ... & TNT \\ 
        58583.3 & 399.2 & 14.888(008) & ... & 14.641(011) & ... & ... & ... & TNT \\ 
        58587.3 & 403.2 & 15.124(009) & 14.853(008) & 14.801(007) & 14.754(009) & ... & 14.718(009) & TNT \\ 
        58590.3 & 406.2 & ... & 15.051(131) & 14.858(116) & ... & ... & ... & TNT \\ 
        58595.3 & 411.2 & 15.548(043) & 15.456(056) & 15.348(054) & 15.261(046) & ... & ... & TNT \\ 
        58604.3 & 420.2 & 16.659(023) & 16.236(011) & 16.337(016) & 16.363(015) & ... & 15.989(012) & TNT \\ 
        58605.3 & 421.2 & 16.592(020) & 16.410(012) & 16.260(015) & 16.211(011) & ... & 16.041(014) & TNT \\ 
        58606.3 & 422.2 & 16.864(031) & 16.560(018) & 16.604(022) & 16.469(015) & ... & 16.350(021) & TNT \\ 
        58607.3 & 423.2 & 16.992(119) & 16.718(116) & 16.639(088) & 16.539(107) & ... & 16.640(183) & TNT \\ 
        58609.3 & 425.2 & 17.496(058) & 17.229(031) & 17.115(039) & 17.026(029) & ... & 16.830(034) & TNT \\ 
        58610.3 & 426.2 & 17.239(032) & 17.150(022) & 16.889(023) & 16.789(017) & ... & 16.657(026) & TNT \\ 
        58611.2 & 427.1 & 17.778(089) & ... & 17.537(065) & ... & ... & ... & TNT \\ 
        58611.3 & 427.2 & ... & 17.699(079) & ... & 17.208(041) & ... & 16.930(036) & TNT \\ 
        58613.2 & 429.1 & 18.653(184) & 18.232(104) & 18.019(097) & 17.502(037) & ... & 17.336(056) & TNT \\ 
        58624.2 & 440.1 & 18.294(446) & 17.748(145) & 17.968(233) & 17.271(098) & ... & 16.971(076) & TNT \\ 
        58631.3 & 447.2 & 18.692(412) & 18.375(168) & 18.068(106) & 17.700(063) & ... & 17.382(068) & TNT \\ 
        58634.3 & 450.2 & 18.740(529) & ... & 18.045(133) & ... & ... & ... & TNT \\ 
        58638.3 & 454.2 & ... & ... & 18.001(086) & ... & ... & ... & TNT \\ 
        58644.2 & 460.1 & 18.573(160) & 18.266(120) & 18.035(176) & 17.806(078) & ... & ... & TNT \\ 
        58644.3 & 460.2 & ... & ... & ... & ... & ... & 17.487(058) & TNT \\ 
        58645.3 & 461.2 & 18.436(732) & ... & ... & ... & ... & ... & TNT \\ 
        58648.2 & 464.1 & ... & 18.174(118) & 17.801(095) & ... & ... & ... & TNT \\ 
        58663.4 & 479.3 & 18.623(030) & ... & 17.954(039) & ... & ... & ... & AZT \\ 
        58664.2 & 480.1 & ... & 18.368(115) & 18.250(182) & 17.781(047) & ... & 17.320(048) & TNT \\ 
        58665.2 & 481.1 & 18.607(187) & 18.272(060) & 18.190(066) & 17.601(037) & ... & 17.179(035) & TNT \\ 
        58669.8 & 485.7 & 18.620(027) & ... & 17.949(031) & ... & ... & ... & AZT \\ 
        58675.3 & 491.2 & 18.612(032) & ... & 17.961(028) & ... & ... & ... & AZT \\ 
        58684.7 & 500.6 & 18.940(028) & ... & 18.183(029) & ... & ... & ... & AZT \\ 
        58708.7 & 524.6 & 15.552(005) & ... & ... & ... & 14.959(003) & ... & Yaoan  \\ 
        58710.7 & 526.6 & 14.836(004) & ... & ... & ... & 14.324(002) & ... & Yaoan  \\ 
        58711.7 & 527.6 & 14.833(004) & ... & ... & ... & 14.299(002) & ... & Yaoan  \\ 
        58720.8 & 536.7 & 14.395(012) & ... & ... & ... & 13.667(003) & ... & Yaoan  \\ 
  \hline
\multicolumn{5}{l}{$^a$ Days relative to the initial optical outburst (UT March 06.58 2018 = MJD 58184.08).}\\
\multicolumn{5}{l}{Note: Uncertainties, in units of 0.001 mag, are $1\sigma$.}\\
\end{tabular}
\end{table*}

\begin{table*}
\centering
\caption{\textit{Swift} Photometry of MAXI J1820+070  \label{sphoto}}
\begin{tabular}{ccccccccc}
\hline
MJD &Phase$^a$ &\textit{uvw2} (mag) & \textit{uvm2} (mag) & \textit{uvw1} (mag) & \textit{UVOT u} (mag)& \textit{UVOT v} (mag)\\
\hline
        58191 & 7 & 12.37(07) & ... & 13.10(06) & 12.52(07) & 12.46(07) \\  
        58192 & 8 & 11.98(09) & 12.18(07) & 12.00(07) & 12.12(07) & 12.63(09) \\  
        58193 & 9 & 11.86(07) & 13.55(11) & ... & 12.10(07) & ... \\  
        58194 & 10 & 11.75(07) & 11.90(09) & 11.88(08) & 12.11(07) & ... \\  
        58195 & 11 & 11.50(07) & 11.84(09) & ... & 11.79(07) & ... \\  
        58197 & 13 & 11.43(07) & 11.55(09) & ... & 11.76(07) & ... \\  
        58198 & 14 & 11.42(07) & 11.58(09) & ... & 11.66(07) & ... \\  
        58199 & 15 & 11.30(09) & 11.48(09) & 11.35(07) & ... & ... \\  
        58201 & 17 & 11.34(09) & 11.49(07) & 11.28(07) & ... & ... \\  
        58202 & 18 & 11.17(09) & 11.28(07) & 11.19(07) & ... & ... \\  
        58204 & 20 & 11.24(09) & 11.34(07) & 11.09(07) & ... & ... \\  
        58206 & 22 & 11.03(09) & 11.15(07) & 11.08(08) & ... & ... \\  
        58208 & 24 & 11.32(09) & ... & 11.32(07) & ... & ... \\  
        58210 & 26 & 11.42(09) & 11.52(07) & 11.29(07) & ... & ... \\  
        58215 & 31 & 11.46(09) & 11.58(07) & 11.48(07) & ... & ... \\  
        58217 & 33 & 11.62(09) & ... & ... & ... & ... \\  
        58219 & 35 & 11.48(09) & ... & ... & ... & ... \\  
        58222 & 38 & 11.59(09) & ... & ... & ... & ... \\  
        58225 & 41 & ... & ... & ... & 12.21(12) & ... \\  
        58226 & 42 & 11.63(09) & 12.39(10) & ... & ... & ... \\  
        58227 & 43 & ... & ... & 11.94(10) & ... & ... \\  
        58229 & 45 & ... & ... & 12.15(10) & 12.40(11) & ... \\  
        58230 & 46 & 12.12(09) & ... & ... & ... & ... \\  
        58231 & 47 & 12.04(09) & ... & ... & ... & ... \\  
        58238 & 54 & 12.13(09) & ... & ... & ... & ... \\  
        58239 & 55 & 11.95(09) & ... & ... & ... & ... \\  
        58241 & 57 & 12.04(09) & ... & ... & 11.95(07) & ... \\  
        58243 & 59 & ... & 12.19(07) & ... & ... & ... \\  
        58244 & 60 & 11.97(09) & ... & ... & ... & ... \\  
        58248 & 64 & ... & ... & ... & 12.19(11) & ... \\  
        58259 & 75 & ... & ... & ... & 12.56(11) & ... \\  
        58259 & 75 & ... & ... & ... & 12.40(10) & ... \\  
        58268 & 84 & 13.33(09) & ... & ... & ... & ... \\  
        58270 & 86 & 12.83(09) & ... & ... & ... & ... \\  
        58272 & 88 & 12.92(09) & ... & ... & ... & ... \\  
        58276 & 92 & 12.87(09) & ... & ... & ... & ... \\  
        58294 & 110 & 12.69(09) & ... & ... & ... & ... \\  
        58306 & 122 & ... & ... & ... & 12.46(10) & ... \\  
        58307 & 123 & 12.38(09) & 12.58(07) & 12.49(07) & 12.50(07) & 13.54(06) \\  
        58308 & 124 & 12.50(09) & 12.54(07) & 12.39(07) & 12.31(07) & 13.46(06) \\  
        58309 & 125 & ... & ... & ... & 12.67(11) & ... \\  
        58310 & 126 & ... & ... & ... & 12.45(11) & ... \\  
        58311 & 127 & 12.17(09) & 12.32(07) & 12.28(07) & 12.17(09) & 13.24(06) \\  
        58312 & 128 & ... & ... & ... & 12.26(11) & ... \\  
        58324 & 140 & ... & 12.58(07) & 12.41(07) & 12.38(08) & 13.43(06) \\  
        58329 & 145 & 12.15(09) & 12.32(07) & 12.19(07) & 12.15(08) & 13.27(06) \\  
        58332 & 148 & 12.14(09) & ... & 12.21(07) & ... & ... \\  
        58334 & 150 & ... & 12.43(07) & 12.25(07) & 12.23(08) & 13.34(06) \\  
        58335 & 151 & 12.22(09) & 12.42(07) & 12.30(07) & 12.25(08) & 13.26(06) \\  
        58337 & 153 & 12.37(09) & 12.48(07) & 12.30(07) & 12.32(08) & 13.35(06) \\  
        58339 & 155 & ... & 12.47(07) & 12.33(07) & 12.32(07) & 13.38(06) \\  
        58363 & 179 & 12.58(09) & ... & 12.51(07) & 12.48(07) & ... \\  
        58369 & 185 & 12.77(09) & 12.91(07) & 12.80(07) & 12.72(07) & 13.76(06) \\  
        58379 & 195 & 12.85(09) & 13.03(07) & 12.94(07) & 12.99(07) & ... \\  
        58383 & 199 & 13.20(09) & 13.28(07) & 13.25(07) & 14.23(06) & 14.27(06) \\  
        58386 & 202 & 13.21(09) & 13.36(07) & 13.22(07) & 13.23(07) & 14.26(06) \\  
        58388 & 204 & 13.22(09) & 13.38(07) & 13.26(07) & 13.28(07) & 14.19(06) \\  
        58390 & 206 & 13.39(09) & 13.56(07) & 13.53(07) & 13.50(07) & 14.52(06) \\  
        58391 & 207 & 13.43(09) & 13.71(07) & 13.55(07) & 13.69(07) & 14.57(06) \\  
        58393 & 209 & 13.49(09) & 13.66(07) & 13.59(07) & 13.73(07) & 14.63(06) \\  
        58395 & 211 & 13.63(09) & 13.77(07) & 13.71(07) & 13.84(07) & 14.82(07) \\  
        58397 & 213 & 13.72(09) & 13.92(07) & 13.79(07) & 13.83(07) & 14.76(07) \\  
        58400 & 216 & 13.54(09) & 13.74(08) & 13.59(07) & 13.62(08) & 14.33(07) \\  

        \hline
\multicolumn{5}{l}{$^a$ Days relative to the initial optical outburst (UT March 06.58 2018 = MJD 58184.08).}\\
\multicolumn{5}{l}{Note: Uncertainties, in units of 0.01 mag, are $1\sigma$.}\\
\end{tabular}
\label{table_lc_uv} 
\end{table*}

\begin{table*}
 \contcaption{A table continued from the previous one.}
 \label{tab:continued}
 \begin{tabular}{ccccccccc}
 \hline
MJD &Phase$^a$ &\textit{uvw2} (mag) & \textit{uvm2} (mag) & \textit{uvw1} (mag) & \textit{UVOT u} (mag)& \textit{UVOT v} (mag)\\
  \hline
        58402 & 218 & 13.54(09) & 13.58(08) & 13.40(07) & 13.41(07) & 14.11(07) \\  
        58404 & 220 & 13.46(09) & 13.60(07) & 13.43(07) & 13.39(07) & 14.07(06) \\  
        58407 & 223 & 13.66(09) & 13.75(07) & 13.65(07) & 13.61(07) & 14.29(06) \\  
        58408 & 224 & 13.62(09) & 13.73(07) & 13.54(07) & 13.62(07) & 14.24(06) \\  
        58410 & 226 & 13.68(09) & 13.77(07) & 13.62(07) & 13.56(07) & 14.35(06) \\  
        58412 & 228 & 13.74(09) & 13.85(07) & 13.69(07) & 13.63(07) & 14.42(06) \\  
        58417 & 233 & 13.94(09) & 14.06(07) & 13.91(07) & 13.91(07) & 14.70(06) \\  
        58420 & 236 & 14.09(09) & 14.15(07) & 14.00(07) & 14.05(07) & 14.73(06) \\  
        58422 & 238 & 14.13(09) & 14.21(07) & 14.11(07) & ... & ... \\  
        58425 & 241 & 14.28(09) & 14.38(08) & 14.26(07) & 14.26(07) & 15.01(07) \\  
        58428 & 244 & 14.35(09) & 14.46(07) & 14.36(07) & 14.35(07) & 15.09(06) \\  
        58431 & 247 & 14.59(09) & 14.62(08) & 14.48(07) & 14.57(07) & 15.23(07) \\  
        58433 & 249 & 14.63(10) & 14.78(09) & 14.49(08) & 14.43(06) & 15.31(09) \\  
        58434 & 250 & 14.83(09) & 14.92(08) & 14.77(07) & 14.70(07) & 15.58(07) \\  
        58437 & 253 & 14.87(09) & 15.04(08) & 14.91(07) & 14.77(07) & 15.53(07) \\  
        58439 & 255 & 15.14(09) & 15.28(08) & 14.94(07) & 14.86(07) & 15.87(07) \\  
        58440 & 256 & 15.21(09) & 15.43(08) & 15.25(08) & 15.21(07) & 15.98(07) \\  
        58546 & 362 & 18.99(15) & 18.81(16) & 18.26(13) & 17.89(12) & 18.15(15) \\  
        58552 & 368 & 17.21(10) & 17.13(10) & 16.57(09) & ... & ... \\  
        58559 & 375 & 13.76(09) & 13.82(07) & 13.68(07) & 13.69(07) & 14.31(06) \\  
        58566 & 382 & 13.44(09) & 13.60(07) & 13.33(07) & 13.41(07) & 14.01(06) \\  
        58576 & 392 & ... & ... & ... & 13.68(07) & ... \\  
        58591 & 407 & ... & 14.58(07) & ... & ... & ... \\  
        58602 & 418 & 15.54(11) & ... & ... & ... & ... \\  
        58607 & 423 & ... & ... & 16.35(07) & ... & ... \\  
        58612 & 428 & ... & ... & 17.65(10) & ... & ... \\  
        58614 & 430 & 18.54(13) & ... & ... & ... & ... \\  
        58615 & 431 & ... & ... & 18.01(20) & 17.57(15) & ... \\  
        58616 & 432 & 18.88(19) & 18.28(19) & 17.97(12) & 17.67(09) & ... \\  
        58620 & 436 & ... & ... & 18.21(11) & ... & ... \\  
        58622 & 438 & ... & 18.75(13) & ... & ... & ... \\  
        58626 & 442 & 18.78(13) & ... & ... & ... & ... \\
        \hline
\multicolumn{5}{l}{$^a$ Days relative to the initial optical outburst (UT March 06.58 2018 = MJD 58184.08).}\\
\multicolumn{5}{l}{Note: Uncertainties, in units of 0.01 mag, are $1\sigma$.}\\
\end{tabular}
\end{table*}

\begin{table*}
\centering
\caption{Log of Spectroscopic Observations of MAXI J1820+070  \label{log} }
\begin{tabular}{ccccccccc}
\hline
MJD &Phase$^{a}$ & Range (\AA) & Disp. (\AA/pix)  & Telescope+Inst.\\
\hline
        58195.93 & 11.85 & 3483-8745 & 2.85 & LJT+YFOSC \\
        58196.86 & 12.78 & 3971-8822 & 2.78 & XLT+BFOSC \\
        58196.93 & 12.85 & 3493-8750 & 2.85 & LJT+YFOSC \\
        58198.92 & 14.84 & 3483-8742 & 2.85 & LJT+YFOSC \\
        58201.81 & 17.73 & 3967-8823 & 2.78 & XLT+BFOSC \\
        58201.92 & 17.84 & 3491-8749 & 2.85 & LJT+YFOSC \\
        58207.83 & 23.75 & 3489-8743 & 2.85 & LJT+YFOSC \\
        58209.85 & 25.77 & 3489-8744 & 2.85 & LJT+YFOSC \\
        58211.82 & 27.74 & 3489-8743 & 2.85 & LJT+YFOSC \\
        58212.83 & 28.75 & 3486-8743 & 2.85 & LJT+YFOSC \\
        58223.86 & 39.78 & 3488-8745 & 2.85 & LJT+YFOSC \\
        58224.76 & 40.68 & 3971-8828 & 2.78 & XLT+BFOSC \\
        58224.76 & 40.68 & 3971-8828 & 2.78 & XLT+BFOSC \\
        58226.82 & 42.74 & 3980-8828 & 2.78 & XLT+BFOSC \\
        58226.83 & 42.75 & 3980-8827 & 2.78 & XLT+BFOSC \\
        58227.78 & 43.70 & 3972-8826 & 2.78 & XLT+BFOSC \\
        58227.80 & 43.72 & 3972-8826 & 2.78 & XLT+BFOSC \\
        58227.82 & 43.74 & 3972-8826 & 2.78 & XLT+BFOSC \\
        58236.74 & 52.66 & 3967-8823 & 2.78 & XLT+BFOSC \\
        58236.76 & 52.68 & 3969-8824 & 2.78 & XLT+BFOSC \\
        58245.81 & 61.73 & 3974-8827 & 2.78 & XLT+BFOSC \\
        58246.82 & 62.74 & 4004-8872 & 2.78 & XLT+BFOSC \\
        58282.75 & 98.67 & 3979-8830 & 2.78 & XLT+BFOSC \\
        58296.61 & 112.53 & 3971-8822 & 2.78 & XLT+BFOSC \\
        58296.63 & 112.55 & 3973-8824 & 2.78 & XLT+BFOSC \\
        58297.66 & 113.59 & 3969-8821 & 2.78 & XLT+BFOSC \\
        58297.68 & 113.60 & 3969-8821 & 2.78 & XLT+BFOSC \\
        58297.69 & 113.61 & 3970-8822 & 2.78 & XLT+BFOSC \\
        58297.71 & 113.63 & 3969-8821 & 2.78 & XLT+BFOSC \\
        58297.72 & 113.64 & 3970-8822 & 2.78 & XLT+BFOSC \\
        58297.74 & 113.66 & 3970-8822 & 2.78 & XLT+BFOSC \\
        58386.55 & 202.47 & 3850-8700 & 2.78 & XLT+BFOSC \\
        58390.46 & 206.38 & 3854-8701 & 2.78 & XLT+BFOSC \\
        58390.48 & 206.40 & 3854-8701 & 2.78 & XLT+BFOSC \\
        58390.50 & 206.42 & 3854-8701 & 2.78 & XLT+BFOSC \\
        58392.46 & 208.38 & 3852-8698 & 2.78 & XLT+BFOSC \\
        58392.48 & 208.40 & 3852-8698 & 2.78 & XLT+BFOSC \\
        58392.50 & 208.42 & 3852-8698 & 2.78 & XLT+BFOSC \\
        58395.46 & 211.38 & 3970-8697 & 2.78 & XLT+BFOSC \\
        58395.48 & 211.40 & 3970-8696 & 2.78 & XLT+BFOSC \\
        58396.49 & 212.41 & 3973-8699 & 2.78 & XLT+BFOSC \\
        58396.50 & 212.42 & 3973-8699 & 2.78 & XLT+BFOSC \\
        58399.44 & 215.36 & 3970-8696 & 2.78 & XLT+BFOSC \\
        58399.46 & 215.38 & 3970-8696 & 2.78 & XLT+BFOSC \\
        58400.45 & 216.37 & 3973-8699 & 2.78 & XLT+BFOSC \\
        58400.47 & 216.39 & 3973-8699 & 2.78 & XLT+BFOSC \\
        58401.45 & 217.37 & 3976-8701 & 2.78 & XLT+BFOSC \\
        58401.47 & 217.39 & 3976-8701 & 2.78 & XLT+BFOSC \\
        58404.48 & 220.40 & 5139-8776 & 4.8 & XLT+OMR \\
        58409.44 & 225.36 & 3762-8689 & 4.8 & XLT+OMR \\
        58421.49 & 237.41 & 3976-8698 & 2.78 & XLT+BFOSC \\
        58431.41 & 247.33 & 3856-8699 & 2.78 & XLT+BFOSC \\
        58435.44 & 251.36 & 3855-8701 & 2.78 & XLT+BFOSC \\
        58441.48 & 257.40 & 3486-8747 & 2.85 & LJT+YFOSC \\
        58444.42 & 260.34 & 3857-8704 & 2.78 & XLT+BFOSC \\
        58446.42 & 262.34 & 3853-8698 & 2.78 & XLT+BFOSC \\
        58447.42 & 263.34 & 3855-8699 & 2.78 & XLT+BFOSC \\
        58555.82 & 371.74 & 4358-8710 & 2.78 & XLT+BFOSC \\
        58559.81 & 375.73 & 3860-8825 & 2.78 & XLT+BFOSC \\
        58559.86 & 375.78 & 3862-8825 & 2.78 & XLT+BFOSC \\
        58564.81 & 380.73 & 3863-8826 & 2.78 & XLT+BFOSC \\
        58564.84 & 380.76 & 3866-8829 & 2.78 & XLT+BFOSC \\
        58566.85 & 382.77 & 3861-8826 & 2.78 & XLT+BFOSC \\
        58568.83 & 384.75 & 3863-8827 & 2.78 & XLT+BFOSC \\
        58568.85 & 384.77 & 3864-8828 & 2.78 & XLT+BFOSC \\
        58568.86 & 384.78 & 3864-8828 & 2.78 & XLT+BFOSC \\
\hline
\multicolumn{5}{l}{$^a$ Days relative to the initial optical outburst (UT March 06.58 2018 = MJD 58184.08).}\\
\end{tabular}
\label{table_log}
\end{table*}


\bsp	
\label{lastpage}
\end{document}